\documentclass[graybox, secnum]{svmult}

\usepackage{undertilde}
\usepackage{amsmath}
\usepackage{amsfonts}
\usepackage{mathptmx}       
\usepackage{helvet}         
\usepackage{courier}        
\usepackage{type1cm}        
\usepackage{makeidx}         
\usepackage{graphicx}        
\usepackage{multicol}        
\usepackage[bottom]{footmisc}
\usepackage{hyperref}        
\usepackage{soul}            
\usepackage{nicefrac}
\hypersetup{colorlinks=true,urlcolor=blue,
	linkcolor = blue,
	urlcolor  = blue,
	citecolor = red,
	anchorcolor = blue  
}
\usepackage[square,numbers]{natbib}
\usepackage{ulem}
\usepackage{cancel}

\newcommand{\nc}{\newcommand}

\nc{\be}{\begin{equation}} 
	\nc{\ee}{\end{equation}} 
\def\bsp#1\esp{\begin{split}#1\end{split}} 
\nc{\beqa}{\begin{eqnarray}} 
	\nc{\eeqa}{\end{eqnarray}} 
\nc{\nn}{\nonumber} 
\nc{\noi}{\noindent} 
\def\bpm{\begin{pmatrix}} 
	\def\epm{\end{pmatrix}} 
\nc{\B}{\Big|}

\nc{\R}{{\cal R}} 
\renewcommand{\E}{{\cal E}}

\nc{\ie}{{\it i.e.}} 
\nc{\eg}{{\it e.g.}} 
\nc{\etc}{{\it etc.}}
\nc{\hc}{{\text{h.c.}}} 
\nc{\e}{\varepsilon}
\nc{\la}{\big<}
\nc{\ra}{\big>} 
\def\g{{\mathfrak g}} 
\renewcommand{\D}{{\cal D}} 
\nc{\tD}{\widetilde{\D}}
\nc{\cD}{\check \D}
\nc{\Dbar}{{\overline \D}}
\nc{\Wbar}{{\overline W}}
\nc{\Dhat}{{\hat D}}
\nc{\Fhat}{{\hat F}}
\nc{\del}{{\partial}} 
\nc{\lag}{{\cal L}}
\nc{\lla}{\langle}
\nc{\rra}{\rangle}
\nc{\EE}{{\cal E}}
\nc{\F}{{ {}^* \hskip -.08truecm F}}
\renewcommand{\d}{{\rm d}} 

\nc{\alphadot}{{\dot\alpha}} 
\nc{\betadot}{{\dot\beta}} 
\nc{\deltadot}{{\dot\delta}} 
\nc{\edot}{{\dot\epsilon}} 
\nc{\gammadot}{{\dot\gamma}} 
\nc{\etadot}{{\dot\eta}} 
\nc{\thetabar}{{\bar\theta}} 
\nc{\Thetabar}{{\bar\Theta}} 
\nc{\Sigmabar}{{\Sigma^\dag}} 
\nc{\psibar}{{\bar\psi}} 
\nc{\etabar}{{\bar\eta}} 
\nc{\ebar}{{\bar\e}} 
\nc{\xibar}{{\bar\xi}} 
\nc{\chibar}{{\bar\chi}} 
\nc{\lambar}{{\bar\lambda}} 
\nc{\sibar}{{\bar\sigma}}

\nc{\is}{{i^\ast}} 
\nc{\js}{{j^\ast}} 
\nc{\ks}{{k^\ast}} 
\nc{\ms}{{m^\ast}} 
\nc{\ls}{{\ell^\ast}} 

\nc{\tm}{{\tilde \mu}} 
\nc{\tn}{{\tilde \nu}} 
\nc{\tr}{{\tilde \rho}} 
\nc{\ta}{{\tilde \alpha}} 
\nc{\tb}{{\tilde \beta}} 
\nc{\taa}{{\widetilde{\dot \alpha}}} 
\nc{\tbb}{{\widetilde{\dot \beta}}} 
\nc{\tM}{{\tilde M}} 
\nc{\bM}{{\bar M}} 
\nc{\tN}{{\tilde N}} 
\nc{\tP}{{\tilde P}} 
\nc{\tQ}{{\tilde Q}}

\nc{\rc}{\textcolor{red}}
\nc{\bc}{\textcolor{blue}}
\nc{\ma}{\textcolor{magenta}}

\def\bpm{\begin{pmatrix}} 
	\def\epm{\end{pmatrix}}

\makeindex             
\makeindex
\begin{document}
	\title*{Supergravity: Application in Particle Physics}
	\author{Florian Domingo  and Michel Rausch de Traubenberg \thanks{corresponding author}}
	\institute{Florian Domingo
		,  Bethe Center for Theoretical Physics \& Physikalisches Institut der Universit\"at Bonn, Nu\ss allee 12, D-53115 Bonn, Germany,   \email{domingo@physik.uni-bonn.de}
		\and Michel Rausch de Traubenberg
		,  Universit\'e de Strasbourg, CNRS, IPHC UMR7178,  23  rue du Loess, Strasbourg, 67037 Cedex, France \email{michel.rausch@iphc.cnrs.fr}}
	%
	%

	\maketitle
	\vskip -3.8truecm
	\abstract{We provide  a pedagogical introduction to   $N=1$ supergravity/super\-sym\-metry in relation to particle physics.  The various steps in the construction of a generic $N=1$ supergravity model are briefly described, and we focus on its low energy supersymmetric limit. The conditions for supersymmetry and supergravity
		breaking are investigated,  and  realistic mechanisms suitable for particle physics identified.
		We then study the model-building aspects  of `softly-broken' supersymmetric 
		extensions of the Standard Model   and discuss several of their phenomenological features.}

\vskip 1.truecm
  \noindent      
{\bf 	
Invited chapter for Handbook of Quantum Gravity (Eds. C. Bambi, L.
Modesto and I.L. Shapiro, Springer Singapore, expected in 2023).}
        \section*{Keywords}
        \vskip -.4truecm
	$N=1$ Supergravity/Supersymmetry, Soft-Supersymmetry Breaking, Gravity-Mediation, Goldstino/Gravitino, MSSM, $R$-Parity, Electroweak Symmetry, Flavour Problem, Dark Matter, Collider Phenomenology, Grand Unification.

	\vskip -5.truecm
	\section{Introduction: Symmetries in Particle Physics}
	Symmetry principles emerge as a fundamental tool in the description of the laws of physics. Since  particle physics is
	usually understood in terms of
	a  Quantum Field Theory (QFT), 
	these principles  enable to classify elementary particles and constrain the form of possible interactions. For instance, elementary particles
	organise according to irreducible representations of the Poincar\'e group  and can be  labelled by the eigenvalues of its two Casimir operators,
	namely their mass and their spin.   Gauge interactions   are associated to a compact Lie group. Spacetime and internal symmetries allow   the construction of multiple  particle physics models;  in particular,
	the so-called \textit{Standard Model} of particle physics,  which seems in capacity  to describe almost all   the currently available high-energy physics data collected in collider experiments.
	
	Since the properties of elementary particles are associated to the underlying  symmetry group  one may wonder what  type  of
	symmetries  are compatible with the basic underlying principles. In this respect QFT places  severe constraints on the possible structures.
	Indeed, if we assume that acceptable  symmetries are encoded
	by Lie algebras, Coleman and Mandula \cite{cm} established that the most
	general form   that the corresponding Lie algebras can take is~\footnote{If we suppose only massless particles a larger symmetry group can be obtained namely the conformal  group.}:
	\beqa
	\g= I\mathfrak{so}(1,3) \times \g_c \nn
	\eeqa 
	where  $I\mathfrak{so}(1,3)$ is the Poincar\'e algebra and $\g_c$ a compact Lie algebra.  All the generators of $\g_c$ commute with those of $I\mathfrak{so}(1,3)$. In other words, all the associated symmetries, in particular those encoding the gauge interactions, are `neutral' with respect to spacetime.
	A fundamental assumption behind the construction of $\g$ is that it contains only  generators
	of a bosonic nature.
	However, due to the spin-statistics and N\oe ther theorems, it is in fact possible to obtain conserved charges of fermionic nature, closing their own algebra though anticommutating relations and thus leading  to  new algebraic structures called  Lie superalgebras.  Within  this latter framework
	it is possible to extend the family of Poincar\'e-compatible symmetries in a non-trivial way ~\cite{hls,gl}, and only supersymmetric extensions~\footnote{If we suppose only massless particles a larger symmetry structure can be obtained namely the superconformal algebra.} of the Poincar\'e algebra are then
	admissible in QFT. The simplest (non-trivial) $N=1$ supersymmetric extension of spacetime symmetries is generated by the  generators of the Poincar\'e algebra $(L_{\mu\nu} = -L_{\nu \mu}, P_\mu, \mu,\nu=0,\cdots,3)$
	and one Majorana spinor $(Q_\alpha, \overline Q_\alphadot, \alpha,\alphadot=1,2)$~\footnote{Following the standard conventions, indices for  left-handed spinors are denoted by $\alpha,\beta,\cdots$, indices for right-handed spinors are taken to be $\dot \alpha,\dot\beta,\cdots$ and vectors indices are given by $\mu,\nu,\cdots$.}.  The non-trivial part of the algebra involving the fermionic charges
	is
	\beqa
	\label{eq:QQ}
	\{Q_\alpha, \overline{Q}_\alphadot\} = -2 \sigma^\mu_{\alpha \alphadot} P_\mu \ ,
	\eeqa
	where $\sigma^\mu_{\alpha \alphadot}$ is defined below --- see after Eq.(\ref{eq:LagSUSY}).
	Thus supersymmetry is a non-trivial extension of the Poincar\'e algebra.
	
	This discovery  subsequently led  to the introduction of  supersymmetry in particle physics.
	Since $Q_\alpha$ is of fermionic nature, supersymmetry is a symmetry  that  maps a boson into a fermion and {\it vice versa}
	and,   as can be proved,   supersymmetric multiplets contain an equal number of bosonic and fermionic degrees of freedom.
	One may also promote the invariance under supersymmetry to a local version: since the anticommutator of two  local supersymmetric transformations amounts to  a spacetime translation (see Eq.[\ref{eq:QQ}]), local supersymmetry  in particular contains gravity,  and is thus called supergravity. 
	
	The purpose of this chapter is to describe applications of supergravity
	in particle physics. To this end, we shall first design a generic theory of fields invariant under $N=1$ supergravity. Then, given that exact supergravity is incompatible with the known particle spectrum, we will study the conditions for supersymmetry and supergravity breaking, and present realistic mechanisms suitable for particle physics: we will more specifically highlight the case of `gravity mediation', where gravitational effects convey supersymmetry breaking from a `hidden' sector to the observable one. Finally, we will discuss several phenomenological features of softly-broken supersymmetric extensions of the Standard Model.

	\section{How to build an action in supergravity}\label{sec:action}
	The first theory of pure supergravity was   obtained  by Freedman, van Nieuwenhuizen \& Ferrara ~\cite{fnf} and Deser \& Zumino  ~\cite{dz} independently.  It took a couple of years
	to understand how matter and and Yang-Mills fields could couple
	to supergravity and  the supergravity coupling of arbitrary interacting matter fields
	was derived in \cite{csj2,csj1,cs1,cfg,bw,ba}.
	The first  attempts  to build invariant Lagrangians  ~\cite{csj2,csj1,cs1,cfg} were performed in  the component formalism, or 
	using  appropriate tensor calculus techniques ~\cite{sw,fn,sw2,sw3,fn2}.
	Then, it was  realised that   a superspace suited  to supergravity
	model building could be introduced
	~\cite{wz,wz2,wz4,gwz2,ss-vas,zum2,sdet,s-space2}.
	The key advantage of this superspace  rests with the automatic invariance under supergravity underlying this formalism, calling all the necessary auxiliary fields and resulting in a suitable action 
	There exist several  books~\cite{wb,west,BK,fp,wein-susy,GGRS,ads6,ads7,freu,mul,nath, fp,mm,rf} and
	reviews~\cite{van,nil,des,bgg} on the topic.

	The purpose of the current section is to provide the reader with the salient points which lead from  general principles  to the four-dimensional $N=1$
	supergravity Lagrangian. The construction closely follows  the book of Wess \& Bagger \cite{wb} and is based on the recent  book and  review by  one of the authors  ~\cite{mm, drv}. In particular these two last references are complementary since
	\cite{mm} offers a pedagogical introduction to supergravity, providing many technical
	details,  whilst technical details are omitted in \cite{drv},  with a focus  on the conceptual scheme that  leads to the four-dimensional $N=1$ supergravity Lagrangian.
	
	\smallskip	
	Four steps can be identified in the derivation of the supergravity-invariant
	Lagrangian.
	The first one  is purely geometrical  and consists in extending   the notion of superspace, already considered in supersymmetry model building ~\cite{ss,ss2,fwz},
	but now in the context of supergravity, {\it i.e.},  constructing  a curved superspace  ~\cite{wz,wz2,wz4, gwz2, BK}. This curved superspace is   such that,  at each superspace point, a tangent space exists, behaving like the traditional flat superspace of global
	supersymmetry. The corresponding structure group is thus  the
	Lorentz group. In particular, in the Einstein frame one has  an invariance under the superdiffeomorphisms whereas in the tangent  or Lorentz frame one has local Lorentz invariance.
	
	Following the standard techniques of general relativity two dynamical variables (which are indeed two superfields)
	are introduced.  The first dynamical superfield is the supervierbein, which connects components in the Einstein frame to  components in  the Lorentz (or flat) frame.  The second dynamical variable is the superconnection, which enables to define covariant derivatives. Then, one associates  (after computing the (anti)commutators of covariant derivatives) two superfields to the two dynamical variables: 
	the torsion and curvature tensors. Some constraints are imposed to the torsion tensor in order to dispose of  overabundant degrees of freedom ~\cite{gwz2,wz2,wz, wb}. These constraints are of physical importance
	since, in addition to drastically reducing the number of degrees of freedom, they allow to explicitly construct physical  models in particle physics that are invariant under supergravity   and  supersymmetric
	transformations  at low energy.
	
	Requiring torsion constraints, the Bianchi identities leads to thirteen equations.
	Solving these equations  allows to express all torsion and curvature tensors in terms of three basic superfields
	~\cite{gz, sie2, wb, dragon}: a chiral
	superfield ${\cal R}$, a chiral symmetric spinor superfield $W_{(\alpha\beta\gamma)}$ and a real vector superfield $G_\mu$.
	It was  observed in \cite{ht} that the structure of constraints upon the torsion tensors is invariant under
	superconformal or Howe-Tucker transformations. Superconformal transformations  are the supergravity analogue of Weyl transformations, or rescaling of the metric, in general relativity and   play a central r\^ole  in the construction of a correctly normalised action in particles physics.
	
	Having  expressed all torsion and curvature tensors in terms of the three basic superfields, and 
	using the large symmetry due to the supergravity algebra, many components can
	be set to zero by means of an appropriate  choice of the parameters  ~\cite{wz, wb,sw}. In particular,  for
	the supervierbein and the superfields ${\cal R}$ and $G_\mu$, we have:
        \index{Supergravity!supervierbein} \index{Gravitino} \index{Supergravity!gravity multiplet!auxiliary fields}
	\beqa \label{eq:E0} 
	E_\tM{}^M(z)\Big| &=& \bpm 
	e_\tm{}^\mu(x) & \frac12 \psi_\tm{}^\alpha(x) & \frac12
	\psibar_{\tm\alphadot}(x) \\ 
	0 & \delta_\ta{}^\alpha & 0 \\ 
	0 & 0 & \delta^\taa{}_\alphadot
	\epm \ , \ \
	\begin{array}{lcl} {\cal R}(z)\Big| &=& -\frac16 M(x) \ , \\
		G_{\mu}(z)\Big|&=& -\frac13 b_{\mu}(x) \ ,
	\end{array}\nn
	\eeqa 
	where  $z \equiv(x,\theta,\bar{\theta})$ corresponds to   a point in a superspace and $X\big|$  represents the lowest order  component of the superfield $X$ in its 
	expansion in terms of the Grassmann variables $\theta, \bar \theta$~\footnote{Lorentz indices in flat space are taken to be untilded $M=(\mu, \alpha, \alphadot)$, whilst 
		Einstein indices in curved space are taken to be tilded $\tM=(\tm, \ta, \taa)$. The three entries are related to the vector, the left-handed spinor  and  the right-handed spinor counterpart of a point in the superspace.} ({\it i.e.}, the fermionic counterpart of the spacetime position $x$ in the superspace). The  fields of the supergravity multiplet are then   the helicity-two
	graviton $ e_\tm{}^ \mu$,     the helicity-$3/2$ gravitino $(\psi_\tm{}^\alpha(x), \psibar_{\tm\alphadot}(x))$ (which is a Majorana spinor-vector) and two auxiliary fields,
	$M$ a complex scalar and $b_\mu$ a real vector. The last field of the gravity multiplet, the connection is a composite  field and can be expressed in terms of the graviton and the gravitino  ~\cite{dz,fnf}.
	
	If we consider a general transformation,  combining an arbitrary superdiffeomorphism and a local Lorentz transformation, there is no reason  why the gauge fixing conditions above  should be preserved. This means that the set of all possible transformations must be restricted
	to the subset protecting the previous gauge condition  ~\cite{wz,wz4, wb}. This restricted set of transformations is called  {\it supergravity
		transformations}. On the one hand a supergravity transformation is  parameterised by a general transformation in the fermionic direction of the superspace, corresponding to a local supersymmetric transformation. On the other hand
	the transformation in the bosonic direction of the superspace and the local Lorentz transformation are  related to  the transformation in the fermionic direction (for the Lorentz transformation it also involves  the auxiliary fields $M$ and $b_\mu$).

	Using  firstly the constraints on the torsion tensors and secondly the
	explicit relationship between the torsion or curvature tensors  with  the supervierbein and the superconnection
	(expressed in the gauge \eqref{eq:E0}) allows to compute the lowest components  of the basic superfields ${\cal R}$, $W_{(\alpha\beta\gamma)}$  and $G_\mu$ together with the lowest
	components of their derivatives with respect to the covariant derivatives \cite{wb}. It turns out that all these lowest components, or the lowest components of derivatives, are expressed in terms of the supergravity multiplet $e_\tm{}^\mu,\psi_\tm{}^\alpha, M,b_\mu$.  \index{Supergravity!gravity multiplet}
	
	Having fixed the gauge, defined what is called  supergravity transformations, and obtained all physical quantities in terms of the supergravity multiplet,  one should then compute the transformation  of the supergravity multiplet under supergravity transformations \cite{wz4,wb,sw,fn,ffnbgo}.
	
	\smallskip
	The second step in the construction consists in introducing superfields in the curved superspace. As in supersymmetry, two types of superfields will be considered in applications of  supergravity in particle physics:  chiral superfields associated to ordinary 
	matter and vector superfields associated to  Yang-Mills theory. It is remarkable  that the
	supersymmetry concepts of chiral and vector superfields  extend to  the supergravity context. This is in fact a consequence of  the torsion constraints, which  amount to an integrability condition.
	Chiral and vector multiplets were independently introduced in supergravity within the tensor calculus
	approach \cite{sw2,sw3,fn2}, the superspace approach \cite{wz4} or when constructing the first invariant Lagrangians
	\cite{fsn,fgsn,ffnbgo,csj1,cs1,csj2,cfg}.
	
	Superfields live  in the curved superspace  but we need to compute  their
covariant derivatives with respect to the
	variables with Lorentz indices. For instance introducing ${\cal D}_\alpha, \overline{{\cal D}}_{\alphadot}$ (pay attention to the fact
	that these are indices in the Lorentz frame and not indices in the Einstein frame) the covariant derivatives with respect to left and right-handed spinors, a chiral superfield is defined by the chirality condition ~\footnote{For the spinor scalar products, we take the usual conventions namely $\chi\cdot\chi = \chi^\alpha \chi_\alpha$ for left-handed spinors and $\bar \chi\cdot \bar \chi = \bar \chi_\alphadot \bar\chi^\alphadot$ for right handed spinors.}  \index{Supergravity!chiral superfield} \index{Supergravity!chiral multiplet} 
	\beqa
	&\overline{{\cal D}}_{\alphadot} \Phi=0 	\label{eq:chir}\\
	&	\varphi=\Phi\B \ , \ \ 
	\chi_\alpha=\frac 1 {\sqrt 2} \D_\alpha \Phi\B\ , \ \ 
	F= \frac14 \D \cdot \D \Phi \B \ . 	\label{eq:Phi}
	\eeqa
	where the second line explicitly provides the components in this gauge.
	They correspond to a scalar field $\varphi$, a left-handed spinor $\chi$ and an auxiliary field $F$.
	A vector superfield is defined by the reality condition $V^\dag = V$. \index{Supergravity!vector superfield} \index{Supergravity!vector multiplet} 
	As in supersymmetry it is possible to select   the Wess-Zumino gauge defined by
	\beqa
	\label{eq:WZ}
	V\Big|=0 \ , \ \ \D_\alpha V\Big|=0\ , \ \ \Dbar_\alphadot V\Big|=0\ , \ \ \D\cdot\D \Phi\Big|=0 \ , \ \ \Dbar\cdot \Dbar \Phi\Big|=0 \ .  
	\eeqa
	For both chiral
	or vector superfields, in order to obtain the full supergravity action, we have to
	compute several (lowest order components of) derivatives.
	This computation is performed using the algebra of supergravity
	(the (anti)commutators of covariant derivatives expressed in terms of the superfields ${\cal R}, W_{(\alpha \beta \gamma)}$ and $G_\mu$). 
	
	At this point,  we would like to stress on a fundamental property  of chiral superfields. If  $\Phi^\dag$  is an anti-chiral superfield then it can be shown that \cite{wz4,gsw}
	$
	\Dbar_\alphadot \big(\Dbar\cdot \Dbar - 8 \R\big) \Phi^\dag = 0 $
	or equivalently  the superfield \index{Supergravity!chiral projector} 
	\beqa
	\label{eq:Xi}
	\Xi= \big(\Dbar\cdot \Dbar - 8 \R\big) \Phi^\dag\ ,
	\eeqa 
	is chiral. 
	This property is central  in obtaining a compact chiral expression of the supergravity action.
	
	Vector superfields can be introduced to describe any Yang-Mills interactions. Consider a compact Lie algebra  $\g_c$  of dimension $n$, 
	a unitary representation $\mathfrak R$ and denote the generators of $\g_c$ in the representation $\mathfrak R$ by
	$T_a=T^\dag_a, \; a=1,\cdots,n$.  If $f_{ab}{}^c \in \mathbb R$ are the structure constants of $\g_c$, the Lie brackets take the form:
	\beqa
	\big[T_a,T_b\big] = i f_{ab}{}^c T_c  \ . \nn
	\eeqa
	Associated to the vector superfield $V= V^a T_a$ we define the spinor superfield strength tensor \index{Supergravity!spinor superfield strength} 
	\beqa
	{\cal W}_\alpha = -\frac 14\big(\Dbar\cdot \Dbar - 8 \R\big) e^{2g V} {\cal D}_\alpha e^{-2gV} \ , \nn
	\eeqa
	where $g$ is the coupling constant.
	Observe that ${\cal W}_\alpha$ is a chiral spinor superfield because of the  `projection' operator $\Dbar\cdot \Dbar - 8 \R$.
	Again  the  algebra of supergravity  enables  to compute (using \eqref{eq:WZ})
	${\cal W}_\alpha\Big|$, ${\cal D}_\beta {\cal W}_\alpha\Big|$ and $\Dbar\cdot \Dbar {\cal W}_\alpha\Big|$.
	It turns out that the degrees of freedom of a  vector superfield  are $v_\mu, (\lambda_\alpha,\bar \lambda_\alphadot)$ and $D$,
	{\it i.e.}, a vector, a Majorana spinor and a real scalar.
	
	We can now assume that the chiral superfield $\Phi$ is in the
	representation  ${\mathfrak R}$ of $\g_c$, {\it i.e.}, transforms like
	$
	\Phi \to e^{-2 gi\Lambda} \Phi \ , \nn
	$
	where $\Lambda = \Lambda^a T_a$ and $\Lambda^a$  are chiral superfields.
	Of course the conjugate of a chiral superfield, which is an anti-chiral  superfield  and whose fermionic component contains a right-handed fermion, is in the representation $\overline{\mathfrak R}$,
	the complex conjugate representation of $\mathfrak R$,  with generators $-T_a^\star=-T_a^t$. It is possible to couple  the chiral superfield $\Phi$ to the vector superfield \cite{wb,mm} in an invariant way by considering 
	$
	{\cal X} =\big(\Dbar \cdot \Dbar - 8 \R\big)\Phi^\dag e^{-2g V}$  (the analogue of \eqref{eq:Xi})
	which  is obviously  chiral, and
	is  again obtained by computing   the lowest order components of its derivatives up to the order two. This computation is
	performed using the  algebra of supergravity, but in the so-called Wess-Zumino gauge for which $V^3=0$.
	
	\smallskip     
	Up to now we have defined the gravity multiplet $(e_\tm{}^\mu, \psi_\tm{}^\alpha, \bar \psi_{\tm \alphadot}, M, b_\mu)$, the matter multiplets $(\varphi^i,\chi^i,F^i), i=1,\cdots, \dim {\cal R}$ and the Yang-Mills multiplets $(v^a_\mu, \lambda^a_\alpha,$ $\bar \lambda^a_\alphadot, D^a),\;  a =1,\cdots, \dim \g_c$: these can finally be combined to construct a supergravity action. \index{Supergravity!gravity multiplet}
        \index{Supergravity!chiral multiplet} \index{Supergravity!vector multiplet}
	As seen in \eqref{eq:Phi}, a chiral superfied is defined in the curved superspace, but
	the constraint \eqref{eq:chir} and its components \eqref{eq:Phi} involve  covariant derivatives with respect to Lorentz indices. This, of course, will make the computation cumbersome. The idea of Julius
	Wess, which can be seen as a {\it tour de force}
	\cite{gp}, consists  in introducing new $\Theta-$variables carrying a Lorentz index instead of an Einstein index, in such a way
	that  the expansion of the chiral superfield $\Phi$ reduces to \cite{wb, wz2}
	\beqa\label{eq:PhiTh}
	\Phi(x, \Theta) = \Phi\B + \sqrt 2 \Theta \cdot (\D \Phi)\B -\frac14
	\Theta\cdot \Theta (\D\cdot \D \Phi)\B
	=
	\varphi + \sqrt{2} \Theta \!\cdot\! \chi - \Theta \!\cdot\! \Theta F \ .\nn
	\eeqa
	
	Next, in general relativity the invariant measure of integration involves the determinant of the vierbein.
	Similarly, in curved superspace the invariant measure of integration involves the superdeterminant of the supervierbein. However, in the context of a chiral form of the Lagrangian (see below) an adapted invariant measure, involving the capacity ${\cal E}$ has to be considered \cite{van,sdet,mull}.
	
	\smallskip
	We  now have all the necessary material to compute the supergravity action, exploiting the notations
	introduced above.
	In the fourth and final step we set $\g_c$ to the specific algebra corresponding to the desired compact Lie group which determines the Yang-Mills interactions. The Lie algebra
	$\g_c$ {can be semisimple (including the product of simple Lie algebras) or can contain some abelian factors or $\mathfrak{u}(1)-$terms. One introduces $\dim \g_c$ vector superfields associated to $\g_c$.  Next,  selecting a matter content in a unitary representation ${\mathfrak R}$ of $\g_c$ (not necessarily
		irreducible), we associate $\dim {\mathfrak R}$ chiral  fields in the representation ${\mathfrak R}$ and $\dim {\mathfrak R}$ anti-chiral  fields in the representation
		$\overline{{\mathfrak R}}$.
		We collectively denote $\Phi^i$ and $\Phi^\dag_\is$ as  chiral or anti-chiral superfields respectively.
		Further, we introduce three
		gauge invariant functions: (1) the superpotential $W(\Phi)$, a holomorphic function depending on chiral superfields;   (2) the gauge kinetic function
		$h_{ab}(\Phi)$ (where $a,b$ are gauge indices in the adjoint representation of $\g_c$),
		a holomorphic function depending on chiral superfields; and (3) the K\"ahler potential  $K(\Phi,\Phi^\dag)$, a real function depending on $\Phi$ and $\Phi^\dag$. The Lagrangian  (in  chiral form) takes the form \cite{ggmw,wb} \index{Supergravity!$N=1$ Lagrangian in superspace}  \index{Supergravity!superpotential} \index{Supergravity!K\"ahler potential}  \index{Supergravity!gauge kinetic function} 
		\beqa \label{eq:Lag} 
		{\cal L}_{\text{SUGRA}} &=& \int   \d^2 \Theta\  \E \Bigg\{\frac{3 m_p^2}8  (\Dbar\!\cdot\!\Dbar -8 {\cal R})  e^{-\frac 1 {3 m_p^2} K(\Phi, \Phi^\dag e^{-2gV})}
		\nn\\
		&&\phantom{xxxxxxxxxx}+ W(\Phi) +   \frac{1}{16 g^2}  h_{ab}(\Phi) {\cal W}^{a \alpha} {\cal W}_\alpha^b\Bigg\}
		+ \text{h.c.} \ ,
		\eeqa
		where $m_p$ is the Planck mass.
		To make it more explicit, we have to expand all fields in the $\Theta-$variables.
		Since all superfields have a large number of
		components, this computation is lengthy,  but not fundamentally  complicated. Once the expansion  has been performed, we have to eliminate the auxiliary fields $F^i,D^a,M,b_\mu$ through their equations of motion, which is  easy as well. At the end of the computation, it turns out that the
		Lagrangian is not canonically normalised. In order to have a canonically normalised Lagrangian we must perform   a dilatation of the
		vierbein, followed by a gravitino shift \cite{wb}, which amount to a redefinition {\it of all the  fields}  and can be interpreted as a superconformal (or Howe-Tucker) transformation. This is certainly the most difficult step.
		
		Due to the technical difficulties
		associated with  the field redefinition, alternative methods have been proposed in order to compute the final Lagrangian.
		The first method \cite{bgg} is geometrical in nature, enlarging the usual superspace to a $U(1)-$superspace.
		The second approach  is based on superconformal methods.
		In  Poincar\'e supergravity, the structure group is the Poincar\'e supergroup. The conformal methods 
		\cite{fktn,ktn,kt,tn,fgn,ft} are based on an enlargement of the structure group to the superconformal group.
		The book of Freedman \& van Proeyen ~\cite{fp} is devoted to conformal methods in supergravity.
		
		The final Lagrangian contains the Einstein-Hilbert action for general relativity, the Rarita-Schwinger Lagrangian for the spin $3/2-$gravitino,
		the kinetic terms of the matter sector (spin $0,1/2$), the kinetic part of the Yang-Mills sector (spin $1,1/2$). All these fields
		are naturally coupled and invariant under supergravity transformations, but also under many other transformations, such as symmetries of the
		underlying K\"ahler manifold (the complex manifold where chiral fields are living),  {\it etc}. Many interacting terms are also generated,
		such as {\it e.g.} four-fermions interactions. The scalar potential takes the form (we  denote $W_i = \partial_i W, K_i= \partial_i K$ {\it etc.},
			$K^\is{}_i =\partial^\is \partial_i K$ the K\"ahler metric of the K\"ahler manifold  and $K^i{}_{i^\ast}$ the inverse  K\"ahler metric)
		~\footnote{In \eqref{eq:scal} we have omitted  the usual $D-$terms (see Sec. \ref{sec:breaking}).}: \index{Supergravity!scalar potential}
		\beqa
		\label{eq:scal}
		V= e^{{\frac 1{m_p^2}K(\varphi,\varphi^\dag)}}\Big({\cal D}_i W {\cal D}^{i^\ast}W^\star  K^i{}_{i^\ast} -\frac 3 {m_p^2} |W|^2 \Big) \ , 
		\nn 
		\eeqa
			where ${\cal D}_i W = W_i +
			1/m_p^2 K_i W$,  with a similar expression for $\overline{\cal D}^\is W^\star$ are covariant derivatives with respect to the K\"ahler manifold structure. It is remarkable  that the structure of the final Lagrangian
		emerges from strictly geometrical and algebraic principles that are related to the properties of the  curved superspace and its associated supergravity transformations. 
		
		\smallskip
		If we now study the low energy limit of \eqref{eq:Lag}, namely when $m_p \to +\infty$, the part of the Lagrangian involving the
		K\"ahler potential becomes:
		\beqa
		{\cal L} = -3 m_p^2  \int   \d^2 \Theta\  \E {\cal R}  -\frac 1 8
		\int   \d^2 \Theta\  \E   (\Dbar\!\cdot\!\Dbar -8 {\cal R})   K(\Phi, \Phi^\dag e^{-2gV}) +\text{h.c.}  + {\cal O}(\frac 1 {m_p^2})\nn \ . 
		\eeqa
		The first piece  reduces to \cite{wb}\index{Supergravity!pure supergravity Lagrangian} \index{Gravitino}
		\beqa
		{\cal L}_{\text{pure~ SUGRA}}=
			e \;m_p^2  \Big( \frac12 R
		+\frac14  \e^{\mu\nu\rho\sigma} \big[ \psi_\mu \sigma_\sigma
		\psibar_{\nu\rho} - \psibar_\mu\sibar_\sigma\psi_{\nu\rho}\big]
		- \frac13  \big[ M M^\star + b_\mu b^\mu \big]\Big)
		\nn
		\eeqa
		where $R$ is the scalar curvature, $e$ is the determinant of the vierbein, $\psi_{\mu \nu}$ is the field strength of the gravitino and $\e^{\mu\nu\rho\sigma}$ is the
		Levi-Civita symbol. This is the Lagrangian obtained in ~\cite{fnf,dz} including the auxiliary fields ~\cite{sw,fn} and describing pure supergravity.
		
		In the limit $m_p \to +\infty$,  the projector $(\Dbar\!\cdot\!\Dbar -8 {\cal R})$ becomes  $\overline{D}\cdot\overline{D}$
		with $\overline D$ (and $D$) the usual covariant derivatives in supersymmetry,  ${\cal W}_\alpha$ becomes $W_\alpha$ the spinor field strength
		in supersymmetry \cite{wb,rf} and $\Theta$ reduce to  $\theta$ the left-handed  Grassmann variables considered in supersymmetry.
		If we now  restrict ourselves to a renormalisable theory the basic functions reduce to:
		\beqa
		\label{eq:renorm}
		K(\Phi,\Phi^\dag)  = \Phi^\dag_{i}  \Phi^i \ ,  \ \ 
		h_{ab}(\Phi)=\delta_{ab}
		W(\Phi)= \alpha_i \Phi^i + \frac 12 m_{ij} \Phi^i \Phi^j + \frac 1 6 \lambda_{ijk}\Phi^i \Phi^j\Phi^k\ . \nn
		\eeqa
		Therefore, omitting Planck-suppressed terms, we obtain in the limit $m_p \to + \infty$\index{Supersymmetry!action in superspace}
		\beqa
		\label{eq:Lagsusy}
		{\cal L}_{\text{SUSY}}= \int \text{d}^2 \theta \Bigg\{-\frac 18 \overline{D} \cdot \overline{D}\Big(\Phi^\dag e^{-2gV} \Phi\Big) \; + \; W+ \frac 1 {16 g^2}\delta_{ab} \; W^{a\alpha} W_\alpha^b \Bigg\}
		+\text{h.c.} 
		\eeqa
		which is the standard action in supersymmetry  expressed in a chiral form. After expanding the fields, we obtain \index{Supersymmetry!action in componants}
		\beqa
		\label{eq:LagSUSY}
		{\cal L}_{\text{SUSY}}=&-&\frac14 \delta_{ab} F^{a \mu \nu}{} F^b_{\mu \nu} -
		\frac{i}{2}\delta_{ab}(\lambda^a \sigma^\mu D_\mu\bar 
		\lambda^b
		-D_\mu \lambda^a\sigma^\mu \bar \lambda_a)+ \frac12\delta_{ab} D^a D^b
		\nn \\
		&+&  
		D_\mu \varphi^\dag D^\mu \varphi -\frac{i}{2}\left(\chi  \sigma^\mu  D_\mu \bar \chi -
		D_\mu \chi \sigma^\mu  \bar \chi \right) +  F^\dag F  \\ 
		&-& g D^a \varphi^\dag T_a{}{} \varphi 
		-i\sqrt{2}g  \bar \lambda^a \cdot \bar \chi T_a \varphi  +i\sqrt{2}g \varphi^\dag \ 
		T_a \chi \cdot  \lambda^a  \nn \\
		&-&\Big(m_{ij}( \varphi^i F^j + \frac12 \chi^i \cdot \chi^j) -
		\frac12\lambda_{ijk}(\varphi^i \varphi^j F^k + \varphi^i \chi^j \cdot \chi^k) + \text{h.c.}\Big)\ , \nn
		\eeqa
		where
		$F_{\mu \nu}^a = \partial_\mu v_\nu^a- \partial_\nu v_\mu^a -g f_{bc}{}^a v_\mu^b v_\nu^c\ , 
		D_\mu \lambda^a = \partial_\mu \lambda^a -g f_{bc}{}^a v_\mu^b \lambda^c$
		are respectively the field strength tensor and the covariant derivative in the Yang-Mills sector, and
		$D_\mu \varphi^i = \partial_\mu \varphi^i + ig (T_a \varphi)^ i \ , 
		D_\mu \chi^i = \partial_\mu \chi^i + ig (T_a \chi)^ i$ 
		are the covariant derivatives in the matter sector. Finally
		$\sigma^\mu_{\alpha \alphadot} = \big(\sigma^0,\sigma^i\big)$ and
		$\bar \sigma^{\mu\alphadot \alpha} = \big(\sigma^0,-\sigma^i\big)$
		where $\sigma^i, i=1,2,3$ are  the usual Pauli matrices and $\sigma^0$ is  the two-by-two identity matrix.  For completeness, we remind our conventions
		for the Minkowski metric and Levi-Civita symbol
		$\eta_{\mu\nu} = \text{diag}\big(1,-1,-1,-1\big)$ and $\epsilon_{0123}=1$.
		If we eliminate the auxiliary fields through their equations of motion  we get
		\beqa
		\label{eq:aux}
		F^\dag_i = \partial_i W \equiv W_i \ \ \text{and} \ \ D_a= g \varphi^\dag T_a \varphi \ 
		\eeqa
		and obtain the scalar potential, with its $F-$ and $D-$terms \index{Supersymmetry!scalar potential}
		part (we denote $D_a = \delta_{ab}D^b, T_a =\delta_{ab} T^b$, {\it etc}.)
		\beqa
		\label{eq:ScalSUSY}
		V&=&V_F +V_D=  F_i^\dag F^i + \frac 12 D_a D^a =  W_i   W^{\star i}  + \frac 12 g^2  (\varphi^\dag T_a\varphi)(\varphi^\dag T^a\varphi)=
		\Big(\alpha_i+ m_{ij} \varphi^i\nn\\
		&&+ \frac 12 \lambda_{ijk}  \varphi^j \varphi^k\Big)
		\Big(\alpha^{\ast i}+  m^{\ast i\ell } \varphi^\dag_\ell  + \frac 12 \lambda^{\ast i\ell m }  \varphi^\dag_\ell \varphi^\dag_m\Big)
		+ \frac 12 g^2 \delta^{ab} (\varphi^\dag T_a\varphi)(\varphi^\dag T_b\varphi)\nn
		\eeqa
		and the Yukawa interactions between spinors and scalar fields \index{Supersymmetry!Yukawa interactions}
		\beqa
		\label{eq:Yuk}
		{\cal L}_{\text{Yuk}} = -\frac 12  W_{ij} \chi^i \cdot \chi^j + \text{h.c.}
		=-\lambda_{ijk} \varphi^i \chi^j \cdot \chi^k + \text{h.c.} 
		\eeqa
		For explicit computations in the supersymmetry context the reader may refer to the book  by one of the author \cite{rf}.
		We further stress that  the supergravity action is derived with all details in \cite{mm,drv}.
		As a complement to  this section one may consult  \cite{DAZ} in the same volume,  where simple supergravity in components is presented, \cite{KuT}  which provides
		a  superspace formulation of supergravity, or \cite{K}, related to Supergravity-Matter couplings in projective superspace.

		\section{Supergravity breaking}\label{sec:breaking}
		
		In the previous section we have described  the basic steps allowing  to construct a Lagrangian coupling matter and Yang-Mills theory to supergravity. Its  limit for    $m_{\text{p}}\to\infty$, {\it i.e.},   the  supersymmetric theory, has  been simultaneously obtained.  This derivation opens the path to applications of the supergravity/supersymmetry formalism to particle physics. At the centre of any such application is the gain in regularity of a QFT protected by supersymmetry, taking the form of non-renormalisation theorems
		\cite{Grisaru:1979wc}. Corresponding particle physics models may then remain perturbative over the `Grand Desert' separating the electroweak from the Planck scales, thus preserving the general structure of the studied model over a huge hierarchy of scales: this is the type of applications in which we specialise below.
		However, this same advantage of regular quantum corrections also finds use in the study of the general properties of Yang-Mills theories at the quantum level (`integrability') \cite{Henn,Dixon,Elvang}.       
		In particular,  implementing  supersymmetry/supergravity in particle physics enables the protection  of the mass of scalar fields from quantum corrections:
		a massless scalar field remains massless at any order of perturbation theory. We will argue in the following paragraph
		that supersymmetry needs to be broken for phenomenological applications; still, this breaking can be enforced in such a `soft' way that this nice 
		property under renormalisation is preserved. In this fashion, softly broken supersymmetry
		provides a technical  solution to the Hierarchy Problem  \cite{wit2,Dimopoulos:1981zb,Witten:1981kv,Kaul:1981hi,Sakai:1981gr}. Nevertheless,
		supersymmetry breaking terms should remain comparable to the electroweak scale so that supersymmetry  actively protects the Higgs mass from comparatively low  scales: observable effects at colliders are then expected.

		Due to the fermionic nature of the generators of supersymmetry/supergravity transformations, a multiplet contains both bosonic and fermionic degrees
		of freedom,  which find a natural embedding in the chiral or the vector superfields of Sec.~\ref{sec:action}.  In particular, if one wants to construct a supersymmetric version of the Glashow-Weinberg-Salam model, each standard particle must be associated to a supersymmetric partner: a scalar to each fermion and a fermion to each boson (gauge and Higgs). Noticing at this point that the operator $P_\mu P^\mu$ is a Casimir operator, particles and their superpartners are necessarily degenerate in mass if supersymmetry is unbroken.  More specifically, this means that if supergravity/supersymmetry were an exact symmetry   of Nature, a massless neutral fermion (associated to the photon) or a light charged scalar (associated to the electron) should have been observed by now. Which is obviously not the case. Thus, supersymmetry/supergravity must be broken at the scale of current particle physics experiments.

		\subsection{Mechanisms of supersymmetry and supergravity breaking} \label{sec:break}
		We first  recall the classical mechanisms achieving a (global) breaking of supersymmetry. One way to  induce a spontaneous breaking consists in designing a field configuration such that the equations of motions for the auxiliary fields $F$ or
		$D$ are incompatible with the trivial solution, {\it i.e.}, at least one of the auxiliary fields develop a vacuum expectation value (vev).  Correspondingly, the scalar potential \eqref{eq:ScalSUSY} becomes strictly positive  at the minimum in a spontaneously broken supersymmetric theory.
		The first mechanism  (O'Raifeartaigh) of supersymmetry breaking achieves $F\neq0$ and involves at least three  chiral superfields \cite{or}. The second mechanism
		(Fayet-Iliopoulos) induces $D\neq0$ through 
		a vector superfield associated to a $U(1)-$gauge symmetry,  allowing for a Fayet-Iliopoulos term in the Lagrangian   \cite{fi,fayet3}.  Both mechanisms fail to produce a spectrum compatible with the experimental situation, as at least some of the `exotic' particles would be lighter than their standard partners.
		Thus, the Standard Model cannot be straightforwardly embedded in a spontaneously broken  supersymmetric framework.  For some details  related to these two mechanisms
		see {\it e.g.} \cite{rf}.

		However,  this picture  could play a  r\^ole  if
		a less simplistic structure is introduced:  supersymmetry is  spontaneously broken in
		a hidden sector   by one of the two mechanisms described above;  then this breaking is transmitted to  an observable (or matter)  sector, which contains a supersymmetric extension of the Standard Model, by the interaction connecting both sectors.
		The nature  of this interaction determines  three classes of  mediation models.
		In the first type, known as gravity-mediated 
		\cite{ca,bfs,iba,ohta,ent,apw,pol}, the supersymmetry breaking is
		communicated to the observable sector {\it via} gravitational interactions. In the second type, called
		gauge-mediation,  supersymmetry is
		broken by a singlet field (called the spurion), which couples ({\it via} the superpotential) to another field, the messenger, belonging to  a non-trivial representation of the gauge group. Quantum loop corrections involving the messenger then break supersymmetry
		in the observable sector \cite{nil,cans1,cans2,dersav,fa-gmsb,df,no,acw,dn1,dn2,dn3,gr,arkani}.  
		In the third type,
		called anomaly mediation,
		the introduction of compensating fields having a conformal (super-Weyl) anomaly induces 
		supersymmetry breaking by purely quantum
		effects \cite{rs,arkani,bmp}.
		Concerning general properties of broken supergravity we refer the reader to
		\cite{wit2,wit, fi,is}.  Here, we will restrict ourselves to the specific case of gravity-induced supersymmetry breaking in Sec.~\ref{sec:gravmed}.
		
		As seen in \eqref{eq:scal} the scalar potential in supergravity, hence its minimum,  is   no longer positive
		(as in global supersymmetry). This turns out to be  an advantage, since, supergravity being  a theory of gravitation,  one expects that  the cosmological constant, hence  the potential at the minimum vanishes.
		Furthermore, as will be  established in Sec. \ref{sec:gold}  after supergravity breaking the gravitino becomes massive.  The assumption of an vanishing cosmological constant, though phenomenologically motivated (as we just stated), appears as a fine tuning problem in general, except in theories  called
		no-scale supergravity (see Sec. \ref{sec:no-scale}) where geometrical properties of the K\"ahler manifold lead to a vanishing potential.  This condition of a vanishing cosmological constant then promotes the gravitino mass  to the status of an order parameter.

		\subsection{The Goldstino}\label{sec:gold}
		\index{Goldstino}
		Let us first consider a supersymmetric theory with the Lagrangian  \eqref{eq:LagSUSY} and suppose that it is broken by some mechanism among those briefly described in  Sec. \ref{sec:break}, {\it i.e.}, that some of the auxiliary fields $F^i$, $D^a$  and some of the scalar fields $\varphi^i$ develop a vev~\footnote{We do not consider fermion condensates in this analysis}. We thus obtain the fermion mass matrix
		(see \eqref{eq:LagSUSY} and \eqref{eq:Yuk})
		\beqa
		{\cal L}_{\text{mass}}=-\frac 12 \bpm \chi^i & i\sqrt{2} \lambda^a\epm
		{\cal M}_F
		\bpm \chi^j \\ i\sqrt 2 \lambda^b \epm\ ,& & {\cal M}_F\equiv \bpm\big<W_{ij}\big> & -g \big<\varphi ^\dag T_b\big>_i\\
		-g \big<\varphi ^\dag T_a\big>_j&0\epm.\nn
		\eeqa
		Using the minimisation of the potential \eqref{eq:ScalSUSY}, the gauge invariance of the superpotential and  \eqref{eq:aux} we have
		
		\begin{equation}
			\begin{cases}
				\partial_i \big<V\big> = \big<W_{ij}\big>\big<F^i\big> + g \big<\varphi^\dag T_a\big>_i \big<D^a\big> =0\\
				\delta_a W^\star =  W^{\star i} \delta \varphi_i^\dag =  -F^i(\varphi^\dag T_a)_i = 0
			\end{cases}\ \ \ \Longrightarrow\ \ \  {\cal M}_F\bpm \big <F^i\big> \\-\big<D^a \big>\epm =0\,,
		\end{equation}
		and thus  a massless field emerges, called the Goldstino or the Goldstone fermion:
		\beqa
		\label{eq:Gold}
		\Psi_G= \frac 1 {\sqrt 2}\Big( \big<F^\dag_i\big> \chi^i -\frac i {\sqrt 2} \big<D_a\big> \lambda^a \Big)=
		\frac 1 {\sqrt 2} \big<F^\dag_i\big> \chi^i -\frac i  2 \big<D_a\big> \lambda^a \ .
		\eeqa
		It is the analogue for broken supersymmetry of the Golstone boson in broken gauge theories   ~\cite{iz, ss3}. However, since such a massless  particle was not observed, again, the absence of a massless fermion in Nature pleads against a spontaneous breaking of global supersymmetry.
		
		The inconvenience of a massless Goldstino is lifted in supergravity by a phenomenon comparable to the Brout-Englet-Higgs mechanism for the electroweak symmetry. We derive it below, assuming for simplicity, and since  this does not modify the conclusion,  a canonical K\"ahler potential and a trivial gauge kinetic function as in \eqref{eq:renorm}. The interaction terms for fermions are the same as in supersymmetry {\it mutatis mutandis} the term $W_{ij}$, which becomes
		${\cal D}_i {\cal D}_j W$ with ${\cal D}_i$, the covariant derivative with respect to the K\"ahler manifold (for notations see \cite{mm}), and the presence of an exponential factor:
		\beqa
		e^{-1}{\cal L}_{\text{ferm}} = -\frac 12 e^{\frac {\varphi^\dag \varphi} {m_p^2}}{\cal D}_i {\cal D}_j W\chi^i \cdot \chi^j
		-i\sqrt{2}g  \bar \lambda^a \cdot \bar \chi T_a \varphi +\text{h.c.}
		\eeqa
		The scalar potential consist of \eqref{eq:scal} (which is in fact the $F-$term and the $N-$term of
		the potential~\footnote{The auxiliary field $N$ is related in a simple way to the field $M$ after an change of variables.}) and the usual $D-$term of supersymmetry. 
		After solving the equation of motion for the auxiliary fields one obtains (see {\it e.g.} \cite{mm}).
		\beqa
		\big<F^i\big> = e^{\frac12 \frac{\big<\varphi^\dag \varphi\big>}{m_p^2} }\big<\overline{{\cal D}}^i W^\star\big>\ , \ \
		\big<D^a\big> = g \big<\varphi^\dag T_a \varphi\big> \ , \ \
		\big<N\big> = -3  e^{\frac12 \frac{\big<\varphi^\dag \varphi\big>}{m_p^2} }\big< W\big>\  \nn
		\eeqa
		and the Golstino of \eqref{eq:Gold} emerges as a massless field, as in supersymmetry.
		
		However, under a supergravity transformation we have
		\beqa
		\delta_\epsilon \chi^i = \sqrt 2 \epsilon F^i + \cdots \ , \ \
		\delta_\epsilon \lambda^a = i \epsilon D^a +\cdots\nn
		\eeqa
		where the dots indicate  terms that are irrelevant for our analysis.  In particular, the Goldstino transforms as:
		\beqa
		\delta_\epsilon \Psi_G = \epsilon\Big(\big<F_i^\dag F^i\big> + \frac 12 \big<D_a D^a\big>\Big) + \cdots
		\eeqa
		which involves the $F-$ and $D-$parts of the scalar potential.
		As the prefactor of the symmetry parameter $\epsilon$ is non-zero for broken supergravity, a local choice of $\epsilon$ allows to purge the Lagrangian from the Goldstino field (similarly to the Goldstone boson in broken gauge theory)
		\cite{csj1,csj2,cfg}.  In this process, the gravitino acquires a mass \index{Gravitino}
		\beqa
		\label{eq:m32}
		m_\frac 32 = \frac1 {m_p^2} \Big<  W e^{\frac 1 2 \frac{K}{m_p^2}}\Big> \ . 
		\eeqa
		The irrelevance of the Goldstino as a dynamical field can also be exhibited by an explicit diagonalisation of the mass matrix, in which a field redefinition makes it disappear ~\cite{wb, fp, mm}.
		Depending on the mediation mechanism, the gravitino mass can take very different values.
		For instance in gauge-mediation it is typically very light, because $m_p-$suppressed, (and it can be the lightest supersymmetric particle). 
		We shall see in the next subsection that in gravity mediation it determines the
		supersymmetry breaking  scale (under  assumption of a vanishing cosmological constant)
		and thus  remains  comparable to  the electroweak scale, {\it i.e.}, $\sim$$100$GeV$-1$TeV.

		\subsection{Gravity induced supersymmetry breaking} \label{sec:gravmed}
                \index{Gravity induced supersymmetry breaking}
		In this section we  focus on gravity-induced supersymmetry breaking.  The chiral superfields are denoted as
		$X^I=(Z^i,\Phi^a),\; i=1,\cdots, m,\;  a=1,\cdots, n$ where $Z$ (resp. $\Phi$) are the superfields in the hidden
		(resp. observable) sector (the corresponding scalars fields are denoted $x^I=(z^i,\varphi^a)$). We do not specify  the
		observable sector here, which  corresponds to a supersymmetric extension of the Standard Model
		(see Sec. \ref{sec:part}).  Of course the model also contains vector superfields associated to the Yang-Mills interactions, but these fields are irrelevant
		for the derivation of the supersymmetry-breaking terms in the  observable sector, which we perform in this subsection.
		In addition, we assume that supergravity is broken in the hidden sector by  an O'Raifeartaigh mechanism.
		More specifically, we suppose that some of the fields of the hidden sector develop a vev of the order of the Planck mass while vevs for the fields in the observable sector
		would be of the order of the electroweak or GUT scales, and  thus   subleading.
		
		Before expanding the scalar potential  \eqref{eq:scal} in the vicinity of its minimum for $m_p \to \infty$
		(now in the presence of vevs of order $m_p$, contrarily to the discussion of Sec.~\ref{sec:action} with unbroken supergravity), it is necessary to specify the behaviour of the terms connecting hidden and observable sectors in this limit: these are assumed to remain finite when $m_p \to \infty$, which considerably restricts the possible form of the K\"ahler potential and superpotential, as was demonstrated  by
		Soni and Weldon in~\cite{sowe}.
		The form retained by Soni and Weldon leads to a soft-breaking of supersymmetry in the observable sector for the limit $m_p\to\infty$, as phenomenologically desirable in traditional applications of supersymmetry to particle physics,
		so that only UV divergences of logarithmic type appear in quantum corrections, thus protecting electroweak scalar masses against corrections from the UV spectrum (Hierarchy Problem). Nevertheless, we note that supergravity breaking admits solutions beyond those of Soni \& Weldon, with in particular possible hard-breaking terms. Models of this type were analysed in \cite{mrt} and are not deprived of phenomenological qualities. Nevertheless, we will focus below on the more commonly studied case of soft-breaking.

		In \cite{gm,bim} the K\"ahler potential and superpotential were chosen as
		\beqa
		K(Z,Z^\dag,\Phi,\Phi^\dag)&=&\widehat{K}(Z,Z^\dag) + \Phi^\dag_{a^\ast}\Lambda^{a^\ast}{}_a(Z,Z^\dag) \Phi^a + \Big(\Gamma_k(Z,Z^\dag)g^k_2(\Phi) +\text{h.c.} \Big)\nn\\
		W(Z,\Phi)&=&\widehat {W}(Z) + W_{ k}(Z) g^k_3(\Phi) \ . \nn
		\eeqa
		Remembering the relative mass dimensions of these quantities
		it is further assumed that
		the realisation of the K\"ahler potential in the hidden (resp.~observable) sector scales like \footnote{This assumption is a consequence of the results of Soni \& Weldon \cite{sw}.}
		$\widehat{ K} \sim m_p^2$ (resp.~$M^2$), while the superpotential is of order $\widehat{ W} \sim M m_p^2$ (resp.~$M^3$), with $M\ll m_p$.
		There are {\it a priori} no restrictions on the form of the functions $g^k_2$ and $g^k_3$ and the general form of the scalar potential after supergravity breaking can be found in \cite{gm,bim}.
		
		For the current analysis,  however, we assume a renormalisable theory in the observable sector, and  specify \index{Supergravity!K\"ahler potential} \index{Supergravity!superpotential}
                 \index{Supergravity!hidden sector}
		\beqa
		\label{eq:KW}
		K(Z,Z^\dag,\Phi,\Phi^\dag)&=&\widehat {K}(Z,Z^\dag) + \Phi^\dag_{a^\ast}\Lambda^{a^\ast}{}_a(Z,Z^\dag) \Phi^a  +
		\Big(\frac 12 \Gamma_{ab}(Z,Z^\dag) \Phi^a \Phi^b + \text{h.c.}\Big) \nn\\
		W(Z,\Phi)&=&\widehat {W}(Z) + \frac 12 m_{ab}(Z) \Phi^a \Phi^b + \frac 16 \lambda_{abc}(Z) \Phi^a \Phi^b \Phi^c \ 
		\eeqa
		{\it i.e.}, the K\"ahler potential (resp.~superpotential) is a quadratic (resp.~cubic) function of the observable fields while the dependence of $\widehat K$ and $\widehat W$ or $\Lambda^{a^\ast}{}_a,\Gamma_{ab},\lambda_{abc}, m_{ab}$ on the hidden fields remains arbitrary.
		Expanding the scalar potential is not intrinsically difficult, but the computation is lengthy because,
		as always in supergravity, there are numerous terms involved. We shall emphasise a few steps of this calculation  (further technical details are given in \cite{mm}).
		Choosing an O'Raifeartaigh breaking mechanism,  at least one
		of the $F^i-$fields  in the hidden sector develops a vev:
		\beqa
		\label{eq:F}
		F^i= \Big<e^{\frac12 \frac {\widehat{ K}}{m_p^2}} \widehat {K}^i{}_{I^\ast} \overline{\D}^{I^\ast} \widehat{ W}^\star\Big> \ne 0 \ , 
		\eeqa
		(note that $\big<K\big> \sim \big<\widehat {K}\big>$ and $\big<W\big> \sim \big<\widehat{ W} \big>$). The order parameter of supersymmetry breaking is then
		given by
		\beqa
		\label{eq:msusy}
		m_{\text{susy}}^4= \big<F^\dag_\js \big> \big<\widehat{K}^\js{}_i\big>\big<F^i\big> \  .
		\eeqa
		\begin{enumerate}
			\item  Inversion of the K\"ahler metric:   Writing
				the K\"ahler  metric and its inverse
			
			\beqa
			K&=& \bpm A&B\\C&D\epm =
			\bpm \partial_{a} \partial^{{a^\ast}}K
			&\partial_{i} \partial^{{a^\ast}}K\\
			\partial_{a} \partial^{{i^\ast}}K&
			\partial_{i} \partial^{{i^\ast}}K\epm   
			\nn \\
			K^{-1}&=& \bpm (A-BD^{-1}C)^{-1}& & -(A-BD^{-1}C)^{-1}BD^{-1}\\
			-D^{-1}C (A-BD^{-1}C)^{-1}& \phantom{ttttt}&D^{-1}C (A-BD^{-1}C)^{-1} BD^{-1} + D^{-1} \epm \ , \nn
			\eeqa
			we  have $A = A_0, B=m^{-1}_p B_1, C= m^{-1}_p C_1, D=m^{-2}_p D_2 + D_0$, with $A_0,B_1,C_1,D_0,D_2$ (not explicitly given here) of order $m_p^0$,    
				so that $D^{-1}= D_0^{-1}(1+m^{-2}_p D_0 D_2)^{-1}$ and $(A-BD^{-1}C)^{-1}=A_0^{-1}(1-m_p^{-2}A_ 0B_1$ $ D^{-1} C_1)^{-1}$ are    perturbatively inverted.
			\item One should compute all covariant derivatives
			$
			\D_I W = \partial_I W + m_p^{-2} W \partial_I K \ , \nn
			$
			and  all terms $\exp(m_p^{-2}K) \D_I W \overline{D}^{J^\ast} W^\star K^I{}_{J^\ast}$ for $(I,J^\ast)=(i,j^\ast),(a,j^\ast), (i,b^\ast)$ and
			$(a,b^\ast)$ corresponding to the hidden-hidden, observable-hidden, hidden-observable, obs\-ervable-observable sectors.
			We stress that these expressions all involve the inverse K\"ahler metric. 
			\item Assume that the scalar fields  $z$ in the hidden  sector are frozen to their vev $\big<z\big> \sim m_p$ and  introduce the gravitino \index{Gravitino}cd
			mass \eqref{eq:m32}. Note that, since  in the expansion of ${\cal D}_IW K^i{}_{J^\ast}{\cal D}^{J^\ast} W^\star -3m_p^{-2} |W|^2$ there are  terms of order $m_p^2$ (see the cosmological constant below in $V_{\text{soft}}$), 
			an additional contribution emerges from the expansion of the exponential factor (the term $1/m_p^2$ coming from the observable sector
			$-$see \eqref{eq:KW}).
			
		\end{enumerate}
		Collecting all terms, the scalar potential takes the form \cite{bim,mm} \index{Softly broken scalar potential}
		\beqa
		\label{eq:Vbreak} 
		V=m_p^2 \Lambda +\partial_a W_m \big<(\Lambda^{-1})^a{}_{a^\ast}\big> \partial^{a^\ast}  W_m^\star  + V_{\text{soft}}   \
		\eeqa
		to which the usual $D-$term (see \eqref{eq:ScalSUSY}) should be added.
		The first term $\Lambda$ (with a summation applying only to the fields of the hidden sector)
		\beqa
		\Lambda= \frac1 {m_p^2} \lla F^\dag_\js\rra \lla\widehat {K}^\js{}_i \rra \lla F^i\rra - 3 |m_\frac32|^2 = \frac 1 {m_p^2} {m_{\text{susy}}^4}-3 \big|m_\frac 32\big|^2 \ , \nn
		\eeqa
		is  the cosmological  constant. \index{Cosmological constant}  It is commonly accepted that, for phenomenologically acceptable theories,
		$\Lambda\approx0$.  Imposing this condition admittedly amounts to considerable fine-tuning. 
		Assuming a vanishing cosmological constant leads to the relation
		\beqa
		\label{eq:scale}
		m_{\text{susy}}^4 = 3 m_p^2 \big| m_\frac 32 \big|^2 \ . 
		\eeqa
		
		The second term of \eqref{eq:Vbreak} is the usual $F-$term of an unbroken supersymmetric theory with superpotential $W_m$ given by 
		\beqa
		\label{eq:Wm}
		&	W_m = \frac16\hat \lambda_{abc} \Phi^a \Phi^b \Phi^c + \frac 12\big( \hat  m_{ab} + m_\frac32 \lla \Gamma_{ab}\rra
		-\big<F^\dag_\is\big> \lla \partial^\is \Gamma_{ab}\rra\big) \Phi^a \Phi^b \ , \\
		&	\hat\lambda_{abc} = e^{{\frac { \lla \widehat{K} \rra}{2 m_p^2}}} \lla\lambda_{abc}\rra \ \ \ \ \text{and}  \ \  \ \ 
		\hat m_{ab}=e^{{\frac  {\lla \widehat {K} \rra}{2 m_p^2} }}\lla m_{ab}\rra \ .	\label{eq:lam}
		\eeqa
		Finally the last term explicitly breaks supersymmetry: \index{Softly broken scalar potential}
		\beqa
		V_{\text{soft}} =   \varphi^\dag_{a^\ast} \Big(\big|m_\frac32\big|^2 S^{a^\ast}{}_a +\Lambda
		\lla\Lambda^{a^\ast}{}_{a}\rra\Big)\varphi^a + \Big(\frac16 A_{abc} \varphi^a \varphi^b\varphi^c  + \frac12 B_{ab} \varphi^a \varphi^b
		+\text{h.c.}\Big) \ , \nn
		\eeqa
		where  \index{Softly broken supersymmetry!trilinear terms}
		\beqa
		\label{eq:A}
		A_{abc}&=&e^{{\frac12 \frac {\lla\widehat {K}\rra}{m_p^2}}}\big<F^i\big>\Bigg[\frac1 {m_p^2}\lla \partial_i \widehat {K} \rra \lla \lambda_{abc} \rra + \lla \partial_i \lambda_{abc} \rra
		\\
		&&-\Big(\lla (\Lambda^{-1})^d{}_{a^\ast} \partial_i \Lambda^{a^\ast}{}_a   \lambda_{dbc}\rra
		+ (a \leftrightarrow b) + (a \leftrightarrow c)\Big)\Bigg] \ , \nn
		\eeqa
		for the trilinear terms \index{Softly broken supersymmetry!bilinear terms}
		\begin{allowdisplaybreaks}
			\beqa
			\label{eq:B}
			B_{ab}&=& e^{{\frac12 \frac {\lla\widehat {K}\rra}{m_p^2}}}\big<F^i\big>\Bigg[\frac1 {m_p^2}\lla \partial_i \widehat {K} \rra \lla m_{ab} \rra + \lla \partial_i m_{ab} \rra 
			-\Big(\lla (\Lambda^{-1})^c{}_{a^\ast} \partial_i \Lambda^{a^\ast}{}_b   m_{ac}\rra + (a \leftrightarrow b) \Big)\Bigg]\nn \\
			&&-m_\frac32^\dag e^{\frac12 \frac {\lla\widehat {K}\rra}{m_p^2}} \lla m_{ab}\rra +(2 |m_\frac32|^2 +\Lambda)\lla \Gamma_{ab}\rra- m^\dag_\frac32 \big<F^\star_\is\big> \lla \partial^\is \Gamma_{ab}\rra
			\\
			&&+m_\frac32 \big<F^i\big> \Bigg[\lla \partial_i \Gamma_{ab} \rra 
			-\Big(\lla (\Lambda^{-1})^c{}_{a^\ast} \partial_i \Lambda^{a^\ast}{}_b   \Gamma_{ac}\rra + (a \leftrightarrow b) \Big)\Bigg]\nn\\
			&& -\big<F^i F^\star_\is\big>\Bigg[\lla \partial_i\partial^\is \Gamma_{ab} \rra 
			-\Big(\lla (\Lambda^{-1})^c{}_{a^\ast} \partial_i \Lambda^{a^\ast}{}_b   \partial^\is \Gamma_{ac}\rra + (a \leftrightarrow b) \Big)\Bigg] \ .,\nn 
			\eeqa
		\end{allowdisplaybreaks}
		for the bilinear terms, and \index{Softly broken supersymmetry!scalar squared masses}
		\beqa
                \label{eq:S}
		S^{a^\ast}{}_a &=&  \lla\Lambda^{a^\ast}{}_a\rra + \frac 1{|m_\frac 32|^2}\lla F^\dag_\is\rra
		\Big<\partial^\is \Lambda^{a^\ast}{}_b \Lambda^{-1})^b{}_{b^\ast}\partial_i \Lambda^{b^\ast}{}_a
		-\partial_i \partial^\is \Lambda^{a^\ast}{}_a\Big>\lla F^i\rra
		\eeqa    
		If \eqref{eq:scale} is satisfied,  $\big<F^i\big> \sim m_p m_\frac32$, using \eqref{eq:m32}, \eqref{eq:F} and the scalings $\big< W \big> \sim \big< \widehat {W} \big> \sim M  m^2_p,\big< K \big> \sim \big< \widehat {K} \big>\sim m_p^2$ and $\big<z\big> \sim m_p$, one arrives at
		\beqa
		\label{eq:rho}
		\big<F^i\big> = m_p m^\dag_\frac32
		\Big<\partial^\js\Big(\frac 1 {m_p} \widehat {K} + m_p \ln \frac {\widehat {W^\star}}{m_p^3}\Big)  \widehat {K}^i{}_\js \Big> = m_p  m^\dag_\frac 32 \hat \rho^i
		\eeqa
		with $\hat \rho^i \sim m_p^{0}$,
		and all soft supersymmetric scales are thus controlled by the gravitino mass, which is then required to be  of the order of the electroweak scale (as an implementation of supersymmetry as a solution of the Hierarchy Problem). \index{Gravitino}
		
		As  announced, the potential $V_{\text{soft}}$ contains terms that break  supersymmetry explicitly, but in a soft way. This means that the loop-quantum corrections associated to  these terms lead  to at most logarithmic ultraviolet divergences. The last soft-breaking term is the gaugino mass
		and is associated to
		a non-trivial gauge kinetic function $h_{ab} \ne \delta_{ab}$ (see {\it e.g.} \cite{mm}). We thus have four types of soft-supersymmetric breaking terms: 
		\begin{itemize}
			\item squared masses for  all the scalar fields (see the $S^{a^\ast}{}_b-$terms);
			\item  masses for all gauginos (not  derived here);
			\item  trilinear couplings among the scalars (see the $A_{abc}-$terms);
			\item  bilinear couplings among the scalars (see the $B_{ab}$-terms). 
		\end{itemize}
		The trilinear terms are directly related to the cubic part of the superpotential whilst the bilinear terms have two origins:
		(1) one related to the quadratic part of the superpotential and (2) one related to the $\Gamma-$term in the K\"ahler potential.  The contribution of the latter type
		is very interesting and  allows to solve the famous $\mu-$problem of the Minimal-Supersymmetric-Standard-Model (see also the $\Gamma_{ab}-$contribution in \eqref{eq:Wm})~\cite{gm}.  (See Sec. \ref{sec:part}.)
		In the presence of an additional linear term $\alpha_a \Phi^a$ in the superpotential (with $\Phi^a$ representing fields of the observable sector that are singlet under all gauge groups), the gravity-mediated supersymmetry breaking mechanism would have generated additional soft-breaking terms of the form $C_a\varphi^a + \text{h.c.}$
		The soft supersymmetric terms were classified in \cite{gg}.
		Terms of a similar form can be generated with the gauge- and anomaly-mediation mechanisms.
		
		At the term of this construction, we recall that the effective K\"ahler potential for the observable fields $\Phi^a$ still reads $K_{\text{eff}}=\Phi_{a^\ast}^{\dagger}\big<\Lambda^{a^\ast}{}_{a}\big>\Phi^a$, so that the kinetic terms for the associated scalars and fermions are not canonically normalised. The last step for an application to particle physics thus amounts to a field re-definition $\Phi^a\to\widehat{\Phi}^i$, enforcing the normalisation $K_{\text{eff}}=\widehat{\Phi}_{i}^{\dagger}\widehat{\Phi}^i$ \cite{mm}. This operation further transforms the effective couplings applying to the $\widehat{\Phi}^i$ basis, as compared to the expressions in Eqs.(\ref{eq:Vbreak}-\ref{eq:S}). In particular, the effective superpotential $W_m(\Phi^a)\to W_m^{\text{eff}}(\widehat{\Phi}^i)$ then contributes to the scalar potential in the canonical form of Eq.(\ref{eq:ScalSUSY}), instead of that of Eq.(\ref{eq:Vbreak}) involving $\big<(\Lambda^{-1})_{a}{}^{a^\ast}\big>$.
		
		As a summary, let us briefly recapitulate the main steps entering the construction of a supergravity model exploitable in particle physics:
		\begin{enumerate}
			\item specify the observable gauge interactions and matter content, respectively described by vector and chiral superfields;
			\item introduce the (gauge invariant) basic functions of Eq.(\ref{eq:KW}) fixing the interplay between observable and hidden sectors;
			\item induce spontaneous supersymmetry breaking in the hidden sector and take the limit $m_p\to\infty$, which results in an effective field theory for the observable sector including soft supersymmetry-breaking terms --- see Eqs.(\ref{eq:Vbreak}-\ref{eq:S});
			\item re-define the fields of the observable sector ({\it via} diagonalisation and rescaling of $\big<\Lambda\big>$) so that their kinetic terms are canonical.
		\end{enumerate}
		We will directly exploit these results in Sect. \ref{sec:part} when setting up a realistic supersym\-metry-inspired model of particle physics, the Minimal Supersymmetric Standard Model.

		\subsection{No-scale supergravity} \label{sec:no-scale}
                \index{Supergravity!no-scale}
		As mentioned in the previous section, gravity-mediated supersymmetry breaking suffers from a fine tuning problem, related to  the cosmological constant, which is artificially set to zero. In \cite{cfkn,ekn,ln}, a class of models, known as no-scale supergravity, was derived, for which the vanishing of the cosmological constant emerges as a property of the underlying K\"ahler manifold where the scalar fields are living. Furthermore, the gravitino mass is generated dynamically. In fact, the scalar potential vanishes identically in such models (one speaks of flat directions in the potential).
		In particular, the gravitino mass is non-vanishing, although classically undetermined, and its values emerges through
		quantum corrections.
		
		Let us  denote as $\Phi^I, I=(0,i), i=1,\cdots, n$ the chiral superfields
		($\Phi^0$ being in the hidden sector and $\Phi^i$ in the observable sector).
		We first notice that it is possible to unify the K\"ahler potential and the superpotential into the so-called
		generalised K\"ahler potential
		\beqa
		{\cal G}(\Phi,\Phi^\dag)= K(\Phi,\Phi^\dag) + m_p^2 \ln \frac{|W|^2}{m_p^6} \ .\nn
		\eeqa
		In fact, this transformation corresponds to a super-conformal transformation, or a K\"ahler transformation, (see {\it e.g.} \cite{mm}).  With this new
		function, the scalar potential takes the form
		\beqa
		V=m_p^2e^{\frac1{m_p^2} {\cal G}}\Big({\cal G}_i {\cal G}^i{}_\is {\cal G}^\is -3 m_p^2\Big) \ ,\nn
		\eeqa
		with ${\cal G}_i =\partial_i {\cal G}, {\cal G}^\is =\partial^\is {\cal G}$ and  ${\cal G}^\is{}_j= \partial_i \partial^\is  {\cal G}$ the K\"ahler metric and
		${\cal G}^i{}_\is$ its inverse.
		
		Choosing
		\beqa
		{\cal G} = - 3 m_p^2\ln\Bigg[\frac{\phi^0 + \phi^\dag_{0^*}}{m^2_p} -\frac{h(\phi^i , \phi^\dag_\is)}{m^2_p}\Bigg]  +m_p^2 F(\Phi^i) +
		m_p^2F^\ast(\Phi^\dag_\is) \nn
		\eeqa
		where  $h$, which contributes to the kinetic part of the scalar $\varphi^i$ and the fermions $\chi^i$,  is an unspecified function and $F$ is a function related to interactions in the observable sector (in fact one can show that $W=m_p^3e^F$ with $W$ the superpotential), a direct computation leads to
		$	{\cal G}_i {\cal G}^i{}_\is {\cal G}^\is =3 m_p^2$  
		Generalised K\"ahler potentials of the above type lead to  no-scale models.
		Such models have been studied  in the  context of SU$(5)$ Grand Unified theories in \cite{ekn,ln} or in the
		context of the Standard Model \cite{sm-no-scale}.

		\section{Supergravity and supersymmetry in particle physics} \label{sec:part} 
		The formalism of the previous sections has provided us with a framework originally embedded in an $N=1$ supergravity theory, but whose observable sector eventually reduces to an $N=1$ supersymmetric model with explicit, albeit soft, supersymmetry-breaking terms. The avowed purpose behind this construction rested with the double wish of, first, embedding the obviously non-supersymmetric spectrum of particle physics (`Standard Model'), at energies comparable to the electroweak scale, second, exploiting the technical protection of scalar masses against quantum corrections from the UV spectrum (Grand Unification, Quantum Gravity, {\it etc.})  in supersymmetric theories. The resulting hybrid indeed amounts to a non-supersymmetric model at energies below the scale of the supersymmetry-breaking terms emerging in the observable sector, $M_{\text{soft}}$ ($\approx m_{\frac32}$ in gravity mediation), but it retains the properties of a supersymmetric theory at energies above that scale. We shall now fulfil our program by explicitly introducing the Standard Model fields in this framework. The reader interested in a detailed discussion of supersymmetric extensions of the Standard Model will read with profit \cite{Drees:2004jm,book_dreiner}.
		\subsection{The Minimal Supersymmetric Standard Model} \label{sec:MSSM}
                \index{MSSM}
		Observing, first, that particle physics at energies comparable to the electroweak scale is well described by the Standard Model, second, that $M_{\text{soft}}$ should be relatively close to the electroweak scale in order to efficiently shield the Higgs squared mass --- radiative corrections still  involve the hierarchy between the electroweak and soft scales ---, it is meaningful to attempt an as-economical-as-possible embedding of the Standard Model in terms of new-physics fields. The product of this operation is known as the Minimal Supersymmetric Standard Model (MSSM)~ \cite{smf1,smf2,f-rep,nil,hk}, and it will be the main model under discussion in this section. Following this principle of minimality, the gauge group remains unchanged as compared to the Standard Model, $G_{\text{SM}}= SU(3)_c\times SU(2)_L \times U(1)_Y$, but its implementation in a supersymmetric context calls for the introduction of the full set of vector superfields, gauge bosons and associated gauginos, in the adjoint representation:
                \index{MSSM!vector multiplets}
		\begin{align}
			\label{eq:V}
			SU(3)_c\,\to&\null\,\hat{G}=(G_{\mu}^a,\tilde{g}^a){\frac{T_a}2}=(\utilde{\bf 8}, \utilde{\bf 1}, 0)=(\text{gluons, gluinos})\nonumber\\
			SU(2)_L\,\to&\null\,\hat{W}=(W_{\mu}^i,\tilde{w}^i){\frac{\sigma_i}2}=(\utilde{\bf 1}, \utilde{\bf 3}, 0)=(W-\text{bosons}, \text{winos})\\
			\phantom{S}U(1)_Y\,\to&\null\,\hat{B}=(B_{\mu},\tilde{b})=(\utilde{\bf 1}, \utilde{\bf 1}, 0)=(B-\text{boson}, \text{bino})\,.\nonumber
		\end{align}
		In the notation $(\utilde{\bf d_3}, \utilde{\bf d_2}, q_1)$, $\utilde{\bf d_3}$, $\utilde{\bf d_2}$ and $q_1$ correspond to the dimension of the $SU(3)_c$ representation, that of the $SU(2)_L$ representation and the hypercharge under which the fields transform. The objects $T_a$ and $\sigma_i$ respectively denote the Gell-Mann and Pauli matrices. Superfields are written with a hat and supersymmetric partners with a tilde.

		Similarly, the fermionic matter content of the Standard model calls for the introduction of three generations ($f=1,2,3$) of chiral superfields describing the quarks and leptons, as well as their supersymmetric counterparts, the squarks and sleptons. Writing only the left-handed fields (the right-handed ones are deduced by hermitian conjugation; the superscript $c$ is just a notation to distinguish
		$SU(2)_L$ singlets from the doublets): \index{MSSM!matter multiplets}
		\begin{align}
			\label{eq:mat}
			& \hat{Q}^f =\Bigg(\bpm \widetilde u^f \\ \widetilde d^f  \epm_L,\bpm u^f \\ d^f  \epm_L\Bigg) =(\utilde{\bf 3}, \utilde{\bf 2}, \phantom{-}\frac16)  \  , & &			\hat{U}^{c\,f}=\Big(\widetilde{u}_R^{c\,f},u_R^{c\,f}\Big)= ( \utilde{\bar{\bf 3}}, \utilde{\bf 1}, -\frac23)  \ ,\nonumber \\
			& & &	\hat{D}^{c\,f}=\Big(\widetilde{d}_R^{c\,f},d_R^{c\,f}\Big) =( \utilde{\bar{\bf 3}}, \utilde{\bf 1},  \phantom{-}\frac13) \ ,  \\
			&\hat{L}^f=\Bigg(\bpm \widetilde \nu^f \\ \widetilde e^f  \epm_L,\bpm \nu^f \\ e^f  \epm_L\Bigg) =
			(\utilde{\bf 1}, \utilde{\bf 2}, -\frac12)  \ , & &
			\hat{E}^{c\,f}=\Big(\widetilde{e}_R^{c\,f},e_R^{c\,f}\Big)= ( \utilde{\bf 1}, \utilde{\bf 1}, \phantom{-}1)\ . \nonumber
		\end{align}
		Finally, the Higgs field requires another chiral supermultiplet $\hat{H} = (\utilde{\bf 1}, \utilde{\bf 2}, -\frac12)$, which it shares with the higgsino. Yet, the above field content would lead to non-vanishing chiral anomalies, hence to a non-renormalisable model, due to the addition of the electroweakly charged higgsino to the fermion sector of the Standard Model. In addition, the holomorphicity of the superpotential forbids the use of conjugate fields when writing the Yukawa terms, so that top and bottom masses cannot be generated with a single Higgs superfield. This prompts for the introduction of two Higgs supermultiplets with opposite hypercharges instead: \index{MSSM!Higgs multiplets}
		\begin{equation}
			\label{eq:Higgs}
			\hat{H}_D =\Bigg(\bpm H^0_{D} \\ H^-_{D}  \epm,\bpm \widetilde{h}_{D}^0 \\ \widetilde{h}_{D}^-  \epm\Bigg) = (\utilde{\bf 1}, \utilde{\bf 2}, -\frac12) \ , \ 
			\hat{H}_U =\Bigg(\bpm H^+_{U} \\ H^0_{U}  \epm,\bpm \widetilde{h}_{U}^+ \\ \widetilde{h}_{U}^0  \epm\Bigg) 
			=(\utilde{\bf 1}, \utilde{\bf 2},\,\frac12) \ .
		\end{equation}
		For the Higgs fields we have further indicated their $U(1)_{\text{em}}-$charges.

		The chiral superfields $\hat{Q}^f$, $\hat{U}^{c\,f}$, $\hat{D}^{c\,f}$, $\hat{L}^f$, $\hat{E}^{c\,f}$, $\hat{H}_U$, $\hat{H}_D$ represent, for the MSSM, the observable fields denoted as $\widehat{\Phi}^i$ at the end of the construction of Sec. \ref{sec:gravmed} --- {\it i.e.},~after enforcing a canonical normalisation of the kinetic terms.
		Following the results of Eqs.(\ref{eq:Vbreak}-\ref{eq:S}) the interactions in the observable sector of the MSSM  can be described by 
		\begin{itemize}
			\item an effective (by assumption renormalisable) superpotential --- we stress that the matrix $\big<(\Lambda^{-1})^a{}_{a^\ast}\big>$ appearing in Eq.(\ref{eq:Vbreak}) becomes trivial after the field re-definition restoring canonical kinetic terms ---, constrained by the symmetries, \index{MSSM!superpotential}
			\begin{align}\label{eq:MSSMW}
				W_{\text{MSSM}}=&\mu \hat{H}_U\cdot \hat{H}_D + Y_u^{ff'}\hat{Q}^f\cdot  \hat{H} \hat{U}^{c\,f'} - Y_d^{ff'}\hat{Q}^f\cdot
				\hat{H}_D \hat{D}^{c\,f'}- Y_e^{ff'}\hat{L}^f\cdot \hat{H}_D  \hat{E}^{c\,f'}\nonumber\\
				&+\mu_i \hat{H}_U \cdot \hat{L}^i+ \lambda_{ijk} \hat{L}^i \cdot \hat{L}^j \hat{E}^{c\,k}
				+ \lambda'_{ijk} \hat{L}^i \cdot \hat{Q}^j \hat{D}^{c\,k} + \lambda''_{ijk} \hat{U}^{c\,i} \hat{D}^{c\,j} \hat{D}^{c\,k}
			\end{align}
			where $\Phi^a\cdot\Phi^b\equiv(\Phi^a)_1(\Phi^b)_2-(\Phi^a)_2(\Phi^b)_1$ stands for the $SU(2)_L$-invariant product (with the indices $1$, $2$ referring to the weak isospin); 
		      \item a set of soft-breaking terms of similar form, with bilinear and trilinear scalar couplings $B$'s and $A$'s as in Eqs.(\ref{eq:A},\ref{eq:B});
                        	\index{Softly broken supersymmetry!bilinear terms} 	\index{Softly broken supersymmetry!trilinear terms} 	\index{Softly broken supersymmetry!scalar squared masses}
			\item squared mass terms $m^{2\,a^*}_{\ \ \ a}\hat \phi^{\dagger}_{a^*}\hat \phi^a$ --- determined by Eq.(\ref{eq:S}) --- where off-diagonal elements only occur for scalar fields with
			identical gauge quantum numbers;
			\item gaugino mass terms $\tfrac{1}{2}M_1\,\tilde{b}\cdot\tilde{b}+\tfrac{1}{2}M_2\,\tilde{w}_i\cdot\tilde{w}_i+\tfrac{1}{2}M_3\,\tilde{g}_a\cdot\tilde{g}_a + \text{h.c.}$
		\index{Softly broken supersymmetry!gaugino masses}\end{itemize}
		Also the $D$-term contribution to the scalar potential should be restored in  Eq.(\ref{eq:Vbreak}). Then, given the arbitrariness of the functions $\widehat{K}$, $\Lambda$ and $\Gamma$ in the hidden sector --- as well as the possibility to call to various breaking mechanisms beyond gravity mediation ---, it is evident from Eqs.(\ref{eq:A}-\ref{eq:S}) that almost any choice of soft-breaking parameters can be {\it a posteriori} justified by a judicious  Ansatz. The simplistic case of trivial functions leads to universality conditions (the flavor index is \index{MSSM!universal softly broken scalar potential}
		omitted):
		\beqa\label{eq:cMSSM}
		V_{\text{soft}}^{{\text{cMSSM}}}&=&
		m_0^2 \Big(|H_U|^2 +  |H_D|^2 +  |Q|^2 + |U^c|^2 + |D^c|^2 + |L|^2 +|E^c|^2\Big)\nn\\
		&& \hskip 1.5truecm + \tfrac{1}{2}M_\frac 12 \Big(\tilde{b}\cdot\tilde{b}+\,\tilde{w}_i\cdot\tilde{w}_i+\,\tilde{g}_a\cdot\tilde{g}_a + \text{h.c.}\Big)\\
		&&   +  \Big[ b \mu {H}_U\cdot {H}_D + a \Big(Y_u{Q}\cdot  {H}_U  {U} - Y_d{Q}\cdot  {H}_D {D}^{c}-  Y_e{L}\cdot {H}_D
		{E}^{c}\Big) + \text{h.c.}\Big]\nn \ . 
		\eeqa
		This choice, (when the second line of \eqref{eq:MSSMW} is not considered) more predictive than realistic,  is only marginally compatible with experimental results today \cite{cMSSM,cMSSM2}.
		
		A discrete symmetry, $R$-parity  \index{R-parity} \cite{Farrar:1978xj}, with charge assignment $1$ to all Standard Model fields and $-1$ to their supersymmetric partners, is often introduced to eliminate the terms of the second line of Eq.(\ref{eq:MSSMW}), as well as the associated soft-breaking couplings. These terms explicitly violate lepton- or baryon-number and lead to a distinctive phenomenology \cite{R-pa}. We discard them till further notice and work with the reduced superpotential $W^{\text{RpC}}_{\text{MSSM}}$ defined by the first line of Eq.(\ref{eq:MSSMW}). An alternative charge assignment, actually  equivalent, is that of a matter parity, with all quark and lepton superfields carrying a charge $-1$, while the Higgs superfields transform trivially.
		
		Let us pause at this point and look at the spectrum in the absence of electroweak-symmetry breaking. Protected by $G_{\text{SM}}$, the quarks, leptons and gauge bosons remain massless, as expected. Squarks and sleptons receive squared masses $\sim M_{\text{soft}}^2$ from the quadratic soft terms: the absence of observed resonances at the Large Hadron Collider tends to push $M_{\text{soft}}$ above the TeV scale. The gauginos take masses of order $M_{\text{soft}}$ from the soft-breaking parameters $M_{1,2,3}$, while higgsinos receive a mass $\mu$ from the superpotential. Several squared-mass scales intervene in the Higgs sector, $\mu^2$ from the superpotential, $B\sim\mu M_{\text{soft}}$ and $m^2_{H_{U,D}}\sim M_{\text{soft}}^2$ from the soft-breaking scalar potential: their interplay must be examined more closely to deduce the conditions of emergence for electroweak-symmetry breaking.

		\subsection{The \boldmath $\mu$-problem}
		It was realised  early on \cite{Kim:1983dt} that the presence of a supersym\-metry-conserving mass term $\mu$ in the superpotential of Eq.(\ref{eq:MSSMW}) was phenomenologically problematic. This parameter being {\it a priori} unrelated to
		supersym\-metry breaking, a natural choice for $\mu$ would involve some high-energy scale, Planck or Grand Unification, much above $M_{\text{soft}}$. However, in such a case, the Higgs potential is dominated by the squared-mass term $|\mu|^2(|H_D|^2+|H_U|^2)$, which no soft contribution can balance, and the electroweak symmetry cannot be broken. Failing to have $\mu$ large, we may set it to $0$, such a choice being protected by the emergence of a $U(1)$ symmetry. Nevertheless, given that the higgsinos take their mass from $\mu$ and that none was observed at the Large Electron-Positron collider, $\mu\geq100$\,GeV, which invalidates this alternative. The non-renormalisation  theorems~\cite{Grisaru:1979wc} also forbid to generate the $\mu$ parameter radiatively. In the aftermath, the unnatural choice $\mu\approx M_{\text{soft}}$ must be retained: this is the $\mu$-problem.
		
		A first solution consists in relating the emergence of $\mu$ to the supersymmetry-breaking mechanism: this is the  proposal by Giudice and Masiero \cite{gm}. In this approach, the $\mu$ parameter is absent in the original superpotential of the supergravity model
		(all parameters are then dimensionless in the K\"ahler potential and the superpotential \eqref{eq:KW}), while
		a term $\Gamma(Z,Z^{\dagger})\hat{H}_U\cdot \hat{H}_D$ appears in the K\"ahler potential --- thus explicitly breaking the $U(1)$ symmetry and avoiding the appearance of an associated Goldstone boson when Higgs fields take a vev. Then the `observable' $\mu$ parameter appears in the limit $m_p\to\infty$ through the $\Gamma$ contributions to the bilinear terms in Eq.(\ref{eq:Wm}), and  is naturally of the order of the gravitino mass. \index{Gravitino} A variant
		\cite{camu} consists in writing a term $\tfrac{1}{m_p^2}\lambda(Z)\widehat{W}(Z)H_U\cdot H_D$ in the superpotential: it can be related to a choice $\Gamma(Z,Z^{\dagger})=\lambda(Z)+\lambda^{\star}(Z^{\dagger})$ by a K\"ahler transformation.
		
		Another solution consists in generating the $\mu$ parameter directly in the observable sector from the vev of an additional singlet superfield $\hat{S}=(\utilde{\bf 1}, \utilde{\bf 1}, 0)$ \cite{fayet3}. The corresponding model is known as the Next-to-Minimal supersymmetric Standard Model (NMSSM) \cite{ers, eht}.  The simplest choice to forbid the $\mu$ term in the superpotential amounts to a  $\mathbb Z_3$ symmetry, allowing only for cubic terms: \index{NMSSM}
		\begin{equation}\label{eq:NMSSMsup}
			W_{\text{NMSSM}}^{{\mathbb Z_3}}=\lambda\hat{S}\hat{H}_U\cdot\hat{H}_D+\frac{\kappa}{3}\hat{S}^3 + \text{Yukawas}\,,
		\end{equation}
		where `Yukawas' stands for the first line in \eqref{eq:MSSMW} without the $\mu-$term.
		Then the minimisation of the scalar potential generally allows $S$ to develop a
		vev $\big<S\big>$, which results in an effective $\mu$ parameter $\mu_{\text{eff}}=\lambda\big<S\big>$. We stress that the only scale entering the scalar potential is the
		supersymmetry-breaking one, thus naturally relating the electroweak-symmetry breaking scale to the latter. In addition, there are numerous phenomenological applications of the new singlet fields, on which we will opportunistically comment in due time. On the other hand, the $\mathbb Z_3$ symmetry proves problematic from the perspective of cosmology, where it causes a Domain-Wall problem
		\cite{Vilenkin:1984ib}. Alternative choices using $R$-symmetries have been studied in \cite{Lee:2011dya}.

		\subsection{Radiative corrections and renormalisation group evolution}
		Meaningful predictions in the MSSM (or any supersymmetric extension of the Standard Model) imply processing this framework as a Quantum Field Theory, {\it i.e.,}~being able to calculate quantum corrections. To this end, it is necessary to regularise ultraviolet divergences appearing in loop diagrams and redefine bare parameters so that such divergences cancel out at the level of observable quantities (renormalisation). The most popular approach to regularisation consists in performing calculations in $D=4-2\varepsilon$ spacetime dimensions instead of $4$. However, this `naive' dimensional regularisation explicitly violates supersymmetry in that vector fields become $4-2\varepsilon$ dimensional while their fermionic counterparts remain $4$-dimensional, hence introducing a mismatch between bosonic and fermionic degrees of freedom. This issue is addressed in dimensional reduction \cite{Siegel:1979wq,Stockinger:2005gx}, where vector fields formally retain $4$ dimensions through the introduction of `epsilon scalars', living in the $2\varepsilon$ dimensions. It is then possible to renormalise the model, {\it e.g.}~through modified minimal subtraction of the ultraviolet divergences, leading to the $\overline{\text{DR}}$ scheme. Alternatively, the $\overline{\text{DR}}'$ scheme \cite{rge5} allows to decouple the epsilon-scalars. We stress that it is not absolutely imperative to work in a framework respecting supersymmetry. One then forfeits the relations between parameters ({\it e.g.}~in the scalar and fermion interactions) that are guaranteed by supersymmetry. Thus, $\overline{\text{MS}}$, for example, remains a legitimate choice. Nevertheless, to ensure that this non-supersymmetric treatment describes a (softly-broken) supersymmetric model, it is ultimately necessary to define its renormalised parameters through their connection to those of a supersymmetry-conserving scheme. \index{Supersymmetry!RGE}  \index{Supersymmetry!radiative corrections} 
		
		In a context involving vastly different scales, such as electroweak physics on one side and gravity-mediated supersymmetry-breaking on the other, it is preferable to resum the ultraviolet logarithms developing between the two scales through the use of the renormalisation group evolution, rather than work with parameters defined at a widely different energy from that of the physical process under study. This coming-and-going between scales is therefore needed, both for testing the predictions of a given high-energy model, {\it e.g.}~of
		supersymmetry breaking, on particle physics, or in view of inferring the structure of the high-energy theory from the low-energy phenomenology. The beta functions of gauge couplings and superpotential parameters are known up to four and three loop, respectively \cite{Jack:1996qq,Jack:1996vg,Jack:1996cn,Jack:1998uj}. On the other hand, the renormalisation group equations for the soft-breaking parameters are generically known at two-loop order \cite{rge1,rge4}, although methods have been proposed and implemented in special cases to include higher orders. As could be anticipated, the running of gauge and superpotential parameters involves only supersymmetry-conserving parameters in their beta functions, while the equations for the soft parameters include both soft-breaking and supersymmetry-conserving parameters. The impact of the running for well separated scales is considerable in general and a set of degenerate soft mass parameters at  $m_p$, as in Eq.(\ref{eq:cMSSM}), --- or at a Grand Unification scale at which several fields are connected by the extended gauge symmetry --- would produce a hierarchical spectrum at the TeV scale.
		
		As an interesting consequence of the modified matter content of the MSSM, with new physics fields in the TeV range, the gauge couplings of $G_{\text{SM}}$ show an approximate convergence when running them up from the electroweak scale towards a unification scale of $M_{\text{GUT}}\approx10^{16}$\,GeV \cite{Amaldi:1991cn}. This feature would allow for a Grand Unification of $G_{\text{SM}}$ in a single step.

		\subsection{Electroweak symmetry breaking and the MSSM Higgs sector}
                \index{Electroweak symmetry}
		The tree-level scalar potential for the Higgs fields in the MSSM reads:\index{MSSM!Higgs potential}
		\begin{multline}\label{eq:Hpot}
			{\cal V}_H=(m^2_{H_D}+|\mu|^2)|H_D|^2+(m^2_{H_U}+|\mu|^2)|H_U|^2+\Big[m^2_{12}\,H_U\cdot H_D+\text{h.c.}\Big]\\+\frac{g_1^2+g_2^2}{8}\Big[|H_D|^4+|H_U|^4\Big]+\frac{g_2^2-g_1^2}{4}|H_D|^2|H_U|^2-\frac{g_2^2}{2}|H_U\cdot H_D|^2
		\end{multline}
		where $g_1$ and $g_2$ are the electroweak gauge couplings, from which one can infer the $D$-term origin of the quartic operators.
		The mass-terms $m^2_{H_{D,U}}\sim M_{\text{soft}}^2$ and $m^2_{12}\sim M_{\text{soft}}\mu$ (substituting the `$B$' notation) denote the quadratic and bilinear soft-breaking terms respectively. Any complex phase in $m^2_{12}$ can be absorbed through a re-definition of the (super)fields, so that the tree-level Higgs potential of the MSSM is automatically CP-conserving --- this is no longer necessarily the case in extended models such as the NMSSM.
		
		The minimisation of Eq.(\ref{eq:Hpot}) --- see the details in {\it e.g.}~\cite{Djouadi:2005gj, rf} --- leads to the solution $(\big<H_D^0\big>,\big<H_U^0\big>)=v(\cos{\beta},\sin{\beta})$ respecting the electromagnetic $U(1)$ symmetry under some conditions among the parameters. The parameter $v$ should be identified to the electroweak symmetry-breaking  vev $(2\sqrt{2}G_F)^{-1/2}$, where $G_F$ is the Fermi constant measured in muon decays. Then, after examining the quadratic terms in the vicinity of this minimum, a triplet of Goldstone bosons $G^0$, $G^{\pm}$ decouples, leaving a CP-odd mass eigenstate $A^0$ with squared mass $M_A^2\equiv\tfrac{2m^2_{12}}{\sin{2\beta}}$ and a charged one $H^{\pm}$, with $M^2_{H^{\pm}}=M_A^2+M_W^2$ and $M_W^2\equiv\tfrac{g_2^2}{2}v^2$. Two states $h^0$ and $H^0$ remain in the CP-even sector and mix according to the mass matrix ($M_{W,Z}$ are the masses of the electroweak gauge bosons):\index{MSSM!Higgs squared mass matrix}
		\begin{equation}\label{eq:CPEmass}
			{\cal M}^2_{\text{CPE}}\equiv\begin{bmatrix}
				M_A^2\,\sin^2\beta+M_Z^2\,\cos^2{\beta} & -(M_A^2+M_Z^2)\sin\beta\cos\beta\\
				-(M_A^2+M_Z^2)\sin\beta\cos\beta & M_A^2\,\cos^2\beta+M_Z^2\,\sin^2{\beta}
			\end{bmatrix}\ \ \ ;\ \ \ M_Z^2\equiv\tfrac{g_1^2+g_2^2}{2}v^2\,.
		\end{equation}
		In this equation the minimisation of the
		Higgs potential has been explicitly applied. The reader may consult \cite{rf} for an explicit computation of the various mass matrix terms.
		
		An effective Standard Model is obtained in the limit $M_A\gg M_Z$, where $h^0$ exactly identifies with the electroweak partner of the Goldstone bosons while $(H^0,A^0,H^{\pm})$ form a heavy (degenerate) doublet. In view of the measured properties of the Higgs particle at $125.25$\,GeV, as well as the phenomenological constraints on non-standard Higgs doublets, this decoupling scenario appears as the most realistic one. Nevertheless, light non-standard Higgs states are still very compatible with collider data as long as they are dominantly singlet, as happens in {\it e.g.}~the NMSSM, due to their suppressed coupling to standard matter.
		
		The lightest eigenvalue of Eq.(\ref{eq:CPEmass}) satisfies $m^2_{h^0}\leq \text{min}(M_Z^2,M_A^2)\cos^2{2\beta}$. Given $M_Z\approx91$\,GeV, $m_{h^0}$ seems much below the observed $125.25$\,GeV. However, this represents no fundamental incompatibility, because the connection between $m_{h^0}$ and $M_Z$ is a tree-level relation enforced by supersymmetry, but, the latter being broken at the electroweak scale, it is not preserved by radiative corrections. The estimated upper bound on the physical mass $M_{h^0}$ before the Higgs discovery amounted to $M_{h^0}\leq135$\,GeV in the MSSM (strongly dependent on the mass of the top quark $m_t$) \cite{Djouadi:2005gj}. Nevertheless, $\tan\beta\gg1$ is favoured in order to maximise the tree-level contribution. Further tree-level effects are possible in extensions of the MSSM, such as a contribution of the $F$-term $\lambda$ --- see Eq.(\ref{eq:NMSSMsup}) --- in the NMSSM, or a mass uplift {\it via} mixing of $h^0$ with a lighter CP-even singlet. Still, radiative corrections to the mass of the Standard-Model-like Higgs reach a considerable relative size, so that a good control of the higher orders is needed to reduce the uncertainties. The corresponding calculations are reviewed in \cite{Slavich:2020zjv}. The leading corrections are controlled by the Yukawa coupling of the top: \index{MSSM!Higgs-mass Quantum correction}
		\begin{equation}
			\Delta m^2_{h^0}\approx \frac{3m_t^4}{4\pi^2 v^2}\Big[\ln\tfrac{M^2_{\tilde{T}}}{m^2_t}+\tfrac{X_t^2}{M^2_{\tilde{T}}}-\tfrac{X_t^4}{12M^4_{\tilde{T}}}\Big]
		\end{equation}
		where $M_{\tilde{T}}\sim M_{\text{soft}}\gg m_t$ represents the average mass of the scalar partners of the top, while $X_t\sim M_{\text{soft}}$ parameterises their mixing, which is induced by electroweak-symmetry breaking. The first term corresponds to an ultraviolet logarithm and shows that, in the presence of a hierarchy between the standard and the supersymmetric sectors, the expansion parameter in the perturbative series is not simply $\tfrac{3m_t^2}{4\pi^2 v^2}$ but $\tfrac{3m_t^2}{4\pi^2 v^2}\ln\tfrac{M^2_{\text{soft}}}{m^2_t}$. This implies the continuing emergence of large effects at higher orders, {\it e.g.}~of the form  $\tfrac{\alpha_Sm_t^2}{\pi^3 v^2}\ln^2\tfrac{M^2_{\text{soft}}}{m^2_t}$, $\tfrac{m_t^4}{\pi^4 v^4}\ln^2\tfrac{M^2_{\text{soft}}}{m^2_t}$ at two-loop. These ultraviolet logarithms can be resummed using the effective field theory techniques, the impact of this resummation being already substantial for $M_{\text{soft}}\approx1$\,TeV. Non-logarithmic and electroweak corrections are naturally also needed for precision predictions. As such, the higher-order uncertainty on the Higgs mass prediction may still remain above $1$\,GeV in scenarios with TeV-scale supersymmetric sectors.
		
		Of course, the Higgs mass is only one electroweak observable among many. For a decoupling scenario, the Higgs couplings are expected to be Standard-Model-like, which is consistent with the current experimental status: narrower determinations in the future may place indirect constraints on the non-Standard spectrum. Leaving the Higgs sector, the relations among the electroweak parameters --- fine structure constant, Fermi constant, $W$- and $Z$-masses and the gauge couplings to fermions --- constrain the non Standard spectrum. The reader may refer to \cite{Heinemeyer:2006px,Heinemeyer:2007bw} for analyses in the MSSM. Nevertheless, for models with only doublet (and singlet)  vevs and a heavy (TeV-like) new-physics sector, such observables are expected to be in good agreement with the Standard Model.
		
		Another precision observable of great interest (though only loosely related to electroweak-symmetry breaking) is the anomalous magnetic moment of the muon, where a historically durable deviation with the Standard Model prediction was recently confirmed by the Fermilab Muon $g-2$ experiment. The discrepancy is of comparable magnitude with the standard electroweak contributions to this observable and could thus hint at new-physics effects at comparable scales. The status of this observable in the MSSM is reviewed in \cite{Stockinger:2006zn}: gauginos, higgsinos and / or sleptons are then expected well below the TeV scale in order to account for the measured anomaly, to which these particles contribute already at one-loop order.
		
		Let us finally comment on the question of the stability of the electroweak-symmetry-breaking vacuum. Indeed, the scalar potential of the MSSM is not restricted to Eq.(\ref{eq:Hpot}), but also includes squarks and sleptons. Vacuum expectation values of charged or coloured fields may lead, for a given choice of Lagrangian parameters, to minima deeper than the electroweak-symmetry-breaking one, thus endangering the phenomenological consistency of an expansion in the vicinity of that specific minimum --- meta-stability may still legitimize the point in parameter space. Approximate analytical constraints on the parameters were derived at tree-level early on. A full analysis of vacuum stability is nevertheless considerably more involved, due to the large dimensionality of the space of scalar fields, and even more so when including radiative corrections. We refer the reader to \cite{Hollik:2018wrr} for a recent discussion and a list of references. 
		
		\subsection{The Supersymmetry flavour problem}
                \index{Flavour problem}
		The flavour structure of the Standard Model is particularly simple in that only the misalignment between the Yukawa coupling matrices in the quark sector allows for flavour transitions governed by the Cabibbo-Kobayashi-Maskawa matrix $V^{\text{CKM}}$ and mediated by the electroweak charged current. This feature is lost when considering a supersymmetric extension of the Standard Model, because the soft-breaking terms in the scalar potential introduce new sources of flavour violation that are potentially independent from the Yukawa couplings: these are the quadratic mass terms $m_{\tilde{F}}^{2\,ff'}$ ($\tilde{F}=\tilde{Q},\tilde{U}^c,\tilde{D}^c,\tilde{L},\tilde{E}^c$) and the trilinear couplings $(A)_{u,d,e}^{ff'}$. Even though an alignment of the soft terms with the Yukawa structure would be engineered at,
		{\it e.g.}, $m_p$ by the supersymmetry-breaking mechanism, this feature is not preserved by the renormalisation group evolution, so that the sfermion couplings at $M_{\text{soft}}$ would still be misaligned with the fermion couplings.
		
		Flavour violation in the quark sector is tested with remarkable accuracy in low-energy transitions such as meson oscillations --- {\it e.g.}~$K-\bar{K}$, $B_d-\bar{B}_d$, $B_s-\bar{B}_s$ --- or rare meson decays --- {\it e.g.}~$K\to\pi\nu\bar{\nu}$, $\bar{B}\to X_s\gamma$, $B_s\to\mu^+\mu^-$, $B^+\to\tau^+\nu_{\tau}$. Flavour violation in the squark sector then typically contributes at the loop level, with squarks and gauginos / higgsinos running in the loop. In particular, the neutral squark interaction with gluinos may induce a flavour transition in association with the strong coupling. If the corresponding flavour structure is completely unrelated to $V^{\text{CKM}}$, experimental bounds on {\it e.g.}~meson oscillation parameters constrain the flavour-violating spectrum up to scales typically reaching beyond $100$\,TeV \cite{Isidori:2010kg}. Slightly milder limits --- the strong coupling does not contribute at leading order --- emerge in the lepton sector from observables such as $\mu\to e\gamma$.
		
		As a consequence of the tight pattern of flavour violation from an experimental perspective, globally consistent with a Standard Model interpretation, one must either renounce new physics close to the TeV scale and set $M_{\text{soft}}\geq100$\,TeV, or assume that the flavour imprint of the Yukawa couplings also applies to the soft supersymmetry-breaking terms at $M_{\text{soft}}\approx1$\,TeV. This latter phenomenological hypothesis is known as Minimal Flavour Violation --- see \cite{Isidori:2010kg} and references therein. In its simplest application, the squark and slepton sectors are exactly aligned with their fermionic counterparts, so that their mass matrices are block-diagonal in the basis of SM flavours and $V^{\text{CKM}}$ is the only source of flavour violation in their charged interactions; flavour transitions can no longer be induced by neutral mediators such as the gluino. Still, flavour observables continue to constrain new-physics effects. The latter, now following the $V^{\text{CKM}}$ structure, are mediated by loops involving charged Higgs and quarks, or charged gauginos / higgsinos and squarks, resulting in limits on the masses of non-standard particles in the TeV range. Small deviations from a strict alignment can of course be considered as well.
		
		This phenomenological picture of alignment at low energy is not really satisfactory from the perspective of model-building and calls for interpretations. The most popular strategy consists in producing the soft supersymmetry-breaking terms {\it via} the gauge-mediation mechanism \cite{Giudice:1998bp}, with comparatively light mediators (`messengers') at about $10-100$\,TeV: gauge interactions would then act in accordance with the existing Yukawa structure in the observable sector and the low scale of the mediation would forbid any significant deviation from alignment to develop {\it via} running effects. The situation is more critical in gravity mediation, where the generation of the soft terms involves physics at $m_p$: a solution to the flavour problem then implies to `guess' the physics behind the flavour structure. A possible approach consists in regarding all flavour structures (Yukawas, soft terms) as spurions, {\it i.e.,}~as relics of fields (`flavons') governing the dynamics behind the breaking of the flavour-symmetry group. The prototype of such constructions is the Froggatt-Nielsen model, using a horizontal $U(1)$ symmetry \cite{Froggatt:1978nt,Grossman:1995hk}.
		
		CP-violation raises an issue comparable to that of flavour to the MSSM. Even under the assumption of a flavour-aligned sfermion structure, several phases of physical meaning (beyond that in $V^{\text{CKM}}$) are  {\it a priori} allowed in the soft sector, in association with the gaugino mass terms $M_{1,2,3}$ and the trilinear couplings $A_{u,d,e}$. On the one hand, new sources of CP-violation are desirable in order to account for the baryon-antibaryon asymmetry of the Universe. On the other, CP-violating phases are tightly constrained by the absence of any experimental evidence for electric dipole moments in nuclei and atoms \cite{Ellis:2008zy}. Constructions similar to those addressing the flavour problem can be designed \cite{Nir:1996am}.
		
		\subsection{Dark Matter phenomenology}
                \index{Dark Matter}
		A distinctive feature of $R$-parity conserving supersymmetry is the stability of the lightest $R$-odd
		({\it i.e.,}~`supersymmetric') particle: indeed, as there exists no lighter final state with $R$-parity $-1$ (by definition), this particle cannot decay if $R$-parity is conserved. Then, according to the usual understanding of the thermal history of the Universe, a stable particle coupled to the Standard Model would leave thermal relics when it drops out of equilibrium in the cooling Universe (`freeze out') \cite{Gondolo:1990dk}. As such, the lightest supersymmetric particle may contribute to the Dark Matter relic density. Given strong astrophysical constraints on the abundance of charged particles, the most promising candidates in the MSSM sector are the (neutral) scalar partners of the neutrinos (sneutrinos) and the neutralinos, {\it i.e.,}~the neutral (and uncoloured) gauginos and higgsinos, which mix after electroweak-symmetry breaking.
		
		All the MSSM Dark-Matter candidates belonging to a non-trivial $SU(2)_L$ multiplet (winos, higgsinos, sneutrinos) tend to annihilate very efficiently in the early Universe, so that they would leave negligible thermal relics unless their mass is in the TeV range or above. On the contrary, a singlet fermion such as the bino --- or the singlino, {\it i.e.,}~the fermionic component of the singlet superfield, in the NMSSM --- tends to lead to excessive relics if left to itself. An excess in Dark Matter is {\it a priori} more problematic than a shortage, as other sectors (beyond the MSSM) could be invoked as complementary (or essential) sources of Dark Matter. However, several mechanisms can be called upon to boost the annihilation cross-sections of singlet fermions: sizable mixing with the other electroweakly charged neutralinos, resonant annihilation in a Higgs (or $Z$) `funnel', presence of a comparatively light $t$-channel mediator, {\it e.g.}~a slepton, existence of a heavier but almost degenerate $R$-odd particle, still abundant at the moment of freeze out and helping in the depletion of $R$-odd particles {\it via} `co-annihilation' processes \cite{Griest:1990kh}. 
		
		The wish to explain all of the observed Dark Matter with a comparatively light MSSM candidate, as well as theoretical prejudice on the spectra, made the case of a bino-dominated Dark matter, dependent on the previous mechanisms, a popular one in the latest few decades. Nevertheless, one should be aware of the numerous caveats behind this hypothesis. First, the lightest $R$-odd particle of the MSSM sector need not be the lightest $R$-odd particle in absolute: obvious competitors would be a lighter gravitino, typically in a gauge-mediation context, or any other exotic particle, {\it e.g.}~an axino, partner of an axion introduced to address the strong CP problem. Then, small $R$-parity violating effects, negligible or not in collider physics, may render the lightest supersymmetric particle unstable on cosmological scales. Finally, non-thermal effects or as yet unknown shortcomings in the formulation of standard cosmology may ruin the thermal picture. Below, we will forget about these warnings and focus on the phenomenology of a weakly-interacting thermal explanation of cold Dark Matter.
		
		We have already copiously discussed the Dark Matter relic density: this quantity is extracted from the measured properties of the Cosmic Microwave Background, corresponding to the radiation trace of the epoch of atom formation (recombination). Its evolution up to this day has been affected through gravitational interactions by the matter / energy content of the Universe. Other evidence for Dark Matter, such as the rotation curves of galaxies or gravitational lensing, can convince us of the existence of Dark Matter in the current Universe, and more specifically in the Solar System. Two main experimental strategies are pursued in the attempt at detecting its by-definition faint interactions with Standard-Model matter. The first one, `Direct Detection', consists in searching for the recoil of heavy nuclei in elastic collisions with Dark Matter particles. Such experiments steadily progress in covering the available plane mass vs.~cross-section down to the `neutrino floor', at which the competition of neutrinos (of Sun, Earth or Cosmic origin) shall raise a challenge for further investigation. The alternative strategy, `Indirect Detection', looks for Dark Matter annihilation currently taking place in regions of space where Dark Matter is expected to be dense, inside massive bodies or near the galactic centre. Typical signals would be energetic neutrinos or gamma-rays, detectable either in Earth-, air- or space-based experiments. In both direct and indirect detection strategies, the identification of a signal involving Dark Matter necessarily depends on the astrophysical modelisation of Dark Matter densities and fluxes in the investigated regions, so that the consequences for particle physics might be tempered. To this day, no robust discovery of a Dark Matter signal in direct or indirect detection has been reported.
		
		Independently of whether the lightest $R$-odd particle of the MSSM represents a sizable component of Dark Matter, one can also attempt to produce it at colliders from Standard Model matter, which leads us to the collider phenomenology. A more detailed overview of supersymmetric Dark Matter may be found in {\it e.g.}~\cite{Catena:2013pka}.
		
		\subsection{Collider phenomenology}
                \index{Collider phenomenology}
		The traditional investigation path in experimental particle physics consists in accelerating electrons and / or protons to very high energies before colliding them: the high kinetic energy of the projectiles may then convert into interactions involving very massive particles, thus providing access to physical effects beyond electromagnetism and nuclear forces. In this fashion, new physics resonances could be directly produced, provided the kinematical threshold is reached and the cross-section $\times$ integrated luminosity is sufficiently large to make such events probable enough in collisions, and detectable with the available experimental sensitivity. An alternative strategy, which we mentioned in previous subsections, consists in precisely measuring the properties of known particle to attempt and detect effects ascribable to physics beyond the Standard Model. We here focus on the direct production approach.
		
		Under the assumption of $R$-parity conservation, one expects the supersymmetric particles to be produced in pairs in collisions of standard matter. The resonances may then either decay to a lighter $R$-odd particle through radiation of standard particles, or escape the detector if they are stable or long-lived. For `usual' MSSM spectra, the strongly or electroweakly interacting $R$-odd particles decay promptly, {\it via} a cascade, down to the lightest supersymmetric particle. In some cases, however, {\it e.g.}~if the decay has very little available phase space, or if couplings to the lightest supersymmetric particle are feeble (or suppressed by the recourse to {\it e.g.}~very massive mediators), a $R$-odd particle may be long-lived at the scales of the collider experiment (where large boost factors can substantially lengthen the apparent lifetime). The most studied scenario is that of a prompt decay into a stable and massive neutralino, which escapes the experiment without being detected, similarly to a neutrino, resulting in a sizable drain of energy and momentum in the apparent balance of the transition. This `missing energy' signature underlies the classical strategy for searches of supersymmetric particles at colliders.
		
		The production of supersymmetric pairs typically occurs through $s$-channel exchange of a gauge boson or $t$- and $u$-channel exchanges of a $R$-odd particle. At lepton colliders, the production process is electroweak, while it can be strong or electroweak at hadron colliders. The expected final states typically involve missing energy, plus jets (strongly interacting matter), plus leptons. A skillful prescription of cuts is usually necessary to distinguish such final states from the Standard Model background, for instance QCD processes at a hadron collider, with radiated $W$'s and $Z$'s producing the leptons and missing energy. The absence of discovery at the Large Electron-Positron collider placed limits on the masses of the supersymmetric particles in the $100$\,GeV range. Similarly, the Large Hadron Collider has been excluding vast areas of the parameter space available to $R$-odd particles, especially those that are produced in strong processes (squarks, gluinos): a typical lower bound on their mass would be one to a few TeV. For uncoloured particles (sleptons, gauginos, higgsinos), the smaller (electroweak) production cross-sections result in milder limits, not exceeding a few $100$\,GeV. Final states with multiple leptons (plus missing energy) are the usual targets in this case. Of course, such experimental constraints are almost never generic, but generally apply to a specific type of spectra, so that their transposition to different scenarios usually requires the recourse to extrapolations or estimates (`recast').
		
		More exotic signatures are also looked for. For instance, long-lived particles decaying on length scales comparable to the detector size may produce leptons or jets with a point of origin distinct from the interaction point (`displaced leptons' and `displaced vertices'). A long-lived (or stable) charged particle could leave identifiable tracks. For larger lifetimes, the deployment of detectors placed at a few $10-100$ meters from the interaction points has been planned at the Large Hadron Collider. However, in the absence of distinctive characteristics such as a large missing energy or the existence of long-lived heavy particles, the identification of new resonances at high energy colliders might prove difficult in general; such a scenario would nevertheless be regarded as highly exotic. Finally, reconstructing the nature of the hypothetically discovered new fields and their classification in a supersymmetric spectrum would call for a long-term, as yet unforeseeable, effort.
		
		The extended Higgs sector of supersymmetric extensions of the Standard Model offers another direction for investigations. The heavy doublet states $H^0$, $A^0$ and $H^{\pm}$ are $R$-even, hence need not be produced in pairs. Typical production modes at the Large Hadron Collider are comparable to those of the Standard Model Higgs boson and involve gluon-gluon fusion, (electroweak) vector boson fusion, assisted production with bottom and / or top quarks. At an $e^+e^-$ collider, these heavy Higgs bosons would have to be produced in pairs or in association with an $h^0$, from a $Z$-boson exchange in the $s$-channel. The dominant decay channels involve fermion pairs, with typical searches in $\tau^+\tau^-$, $\mu^+\mu^-$, $b\bar{b}$ or $t\bar{t}$ ($\tau\bar{\nu}_{\tau}$ or $b\bar{t}$ for $H^-$). Bosonic decay channels are suppressed in the MSSM, although $H^0\to h^0h^0$ may be detectable, provided $H^0$ is light enough (though with mass above threshold). In the NMSSM, however, Higgs-to-Higgs cascade decays involving singlet states may be dominant and overshadow the fermionic modes: these processes are also actively looked for.
		
		A short summary of the search for supersymmetry at the Large Hadron Collider can be found in \cite{Canepa:2019hph}.
		
		\subsection{Supersymmetric Grand Unification}
		
		The MSSM or its singlet extension (NMSSM) are the simplest models fulfilling a softly-broken supersymmetric embedding of the Standard Model. However, about any idea in non-supersymmetric particle physics, {\it e.g.}~axions, vector quarks, right-handed neutrinos, can be transposed to the supersymmetry /  supergravity-breaking framework of section \ref{sec:breaking}. Here we shall provide a few insights concerning Grand Unification in this context. Two coincidences of the Standard Model (or the MSSM) particularly motivate the idea of a unification of gauge interactions within a  less disparate group. The first one is the cancellation of the chiral anomalies, accidentally resulting from the matter content. The second one is the quantisation of the electric charge, {\it i.e.,}~the fact that all elementary particles take charges that are multiple of $1/3$: the hypercharge being a $U(1)$ symmetry, there is no deep reason why the various hypercharges in the Standard Model should appear in rational proportions. Reciprocally, the protection of radiative corrections by
		supersymmetry benefits grand unified constructions through the stabilisation of scale hierarchies and the reduced risk of encountering a non-perturbative regime (Landau pole). Finally, we emphasise that the apparent convergence of the Standard Model gauge couplings at $M_{\text{GUT}}\approx10^{16}$\,GeV in the MSSM favours a one-step unification ($SU(5)$, $SO(10)$, {\it etc.}), with a `Grand Desert' between the electroweak and Grand Unification scales, rather than the multistep path of left-right symmetry and Pati-Salam. For reviews of the group theoretical concepts involved in Grand Unification, we refer the reader to ({\it e.g.})~\cite{Slansky:1981yr,Ross:1985ai,Yamatsu:2015npn, rm}, and to \cite{Mohapatra:1999vv,Mohapatra_book,Raby:2017ucc} concerning their application in a supersymmetric context.
		
		The simplest embedding of $G_{\text{SM}}$ in a compact simple Lie group employs an $SU(5)$ gauge symmetry, thus
		involving no reduction of rank in spontaneous symmetry breaking from $SU(5)$  to $G_{\text{SM}}$.
		One may view the first three rows of the fundamental $SU(5)$ representation as transforming under the
		fundamental representation of $SU(3)_c$
		and the two last, under the fundamental representation of $SU(2)_L$.
		Finally, the generator ${\bf t^{24}}=\sqrt{\tfrac{3}{5}}\text{diag}\big(-\tfrac{1}{3},-\tfrac{1}{3},-\tfrac{1}{3},\tfrac{1}{2},\tfrac{1}{2}\big)$ commutes with all $SU(3)_c$ and $SU(2)_L$ generators and can be identified, up to a proportionality constant, to the hypercharge: ${\bf Y}=C^Y_{24}{\bf t^{24}}$. The conventional normalisation of the hypercharge leads to $C^Y_{24}=\sqrt{\tfrac{5}{3}}$
		(because all generators  in the fundamental representation satisfy
		Tr$({\bf t}^a {\bf t}^b) = \nicefrac {\delta^{ab}} {2}$). This connection explains the quantisation of the electric charge. Then, studying the branching rules of $SU(5) \supset G_{\text{SM}}$, one observes:
		\begin{align}
			\utilde{\overline{\bf 5}}&=(\utilde{\bf \bar{3}}, \utilde{\bf 1}, \tfrac{1}{3})\oplus(\utilde{\bf 1}, \utilde{\bf 2}, -\tfrac{1}{2})\,,\hspace{1cm} \utilde{{\bf 5}} =(\utilde{\bf 3}, \utilde{\bf 1}, -\tfrac{1}{3})\oplus(\utilde{\bf 1}, \utilde{\bf 2}, \tfrac{1}{2})\,,\nonumber\\ 
			\utilde{\bf 10}&=(\utilde{\bf \bar{3}}, \utilde{\bf 1}, -\tfrac{2}{3})\oplus(\utilde{\bf 1}, \utilde{\bf 1}, 1)\oplus(\utilde{\bf 3}, \utilde{\bf 2}, \tfrac{1}{6})\,,\\
			{\bf 24}&=(\utilde{\bf 8}, \utilde{\bf 1}, 0)\oplus(\utilde{\bf 1}, \utilde{\bf 3}, 0)\oplus(\utilde{\bf 1}, \utilde{\bf 1}, 0)\oplus(\utilde{\bf 3}, \utilde{\bf 2}, -\tfrac{5}{6})\oplus(\utilde{\bf \bar{3}}, \utilde{\bf 2}, \tfrac{5}{6})\,.\nonumber	
		\end{align}
		Consequently, the chiral superfields containing the Standard model fermions can be embedded in (three generations of) a
		$\utilde{\overline{\bf 5}}^f_F\oplus\utilde{{\bf 10}}^f_F$ representation of $SU(5)$ (see \eqref{eq:mat}), while $H_{U,D}$ are identified with the doublet components of $\utilde{\bf 5}_S\oplus\utilde{\overline{\bf 5}}_S$ (see \eqref{eq:Higgs}). In the adjoint representation ($\utilde{\bf 24}$) of $SU(5)$, one finds an embedding for all the Standard Model gauge (super)fields, with the additional $(\utilde{\bf 3}, \utilde{\bf 2}, -\tfrac{5}{6})\oplus(\utilde{\bf \bar{3}}, \utilde{\bf 2}, \tfrac{5}{6})$ called lepto-quarks. The latter take mass at the scale of the breaking $SU(5)\to G_{\text{SM}}$ and convey new physics effects, such as lepton- and baryon-number violation. These should remain small at the electroweak scale for a phenomenologically realistic model, so that the breaking scale needs satisfy $M_{\text{GUT}}\gg M_Z$. In fact $M_{\text{GUT}}$ is most naturally chosen as the scale at which the standard gauge couplings converge, {\it i.e.,}~$M_{\text{GUT}}\approx10^{16}$\,GeV for a MSSM field content with $M_{\text{soft}}\approx 1$\,TeV. The field content described above does not allow the breaking $SU(5)\to G_{\text{SM}}$ ({\it i.e.}, ~would break $G_{\text{SM}}$ simultaneously with $SU(5)$), so that, as generic in grand unified models, further fields belonging to representations of larger dimensions are needed: the simplest choice consists in introducing $\utilde{\bf 24}_S$ in the adjoint representation of $SU(5)$ and taking a
		vev proportional to ${\bf t^{24}}$. This field content is again (accidentally) anomaly-free.
		
		We may now write the superpotential for the $SU(5)$ model thus designed:\index{GUT!$SU(5)$}
		\begin{multline}\label{eq:SU5W}
			W_{SU(5)}=\frac{m}{2}\,\text{Tr}(\utilde{\bf 24}_S)^2+\frac{\lambda}{3}\,\text{Tr}(\utilde{\bf 24}_S)^3+\mu(\utilde{\bf 5}_{S})_i(\utilde{\overline{\bf 5}}_S)^i+\beta(\utilde{\overline{\bf 5}}_S)^i(\utilde{\bf 24}_S)_{i}^{\ j}(\utilde{\bf 5}_{S})_j\\+Y_5(\utilde{\overline{\bf 5}}_F)^i(\utilde{\bf 10}_{F})_{ij}(\utilde{\overline{\bf 5}}_S)^j+Y_{10}\varepsilon^{ijk\ell m}(\utilde{\bf 10}_{F})_{ij}(
			\utilde{\bf 10}_{F})_{k\ell}(\utilde{\bf 5}_{S})_{m}
		\end{multline}
		where the indices $i,\ldots,m$  correspond to the transformation under the (anti)funda\-mental $SU(5)$ --- the generation indices are omitted ---and $\varepsilon^{ijk\ell m}$ is the five-dimensional Levy-Civita symbol. The matter-parity, {\it i.e.,}~invariance under
		$\utilde{\overline{\bf 5}}_F,\utilde{\bf 10}_F\to-\utilde{ { \overline{\bf 5}}}_F,-{\bf 10}_F$, has been implicitly required.
		The $SU(5)$-breaking minimum $<\utilde{\bf 24}_S>=\tfrac{m}{\lambda}\text{diag}(2,2,2,$ $-3,-3)$ does not break
		supersymmetry and is therefore degenerate with the $SU(5)$-conserving one as long as
		supersym\-metry-breaking terms are not introduced. The symmetry breaking $G_{\text{SM}}\to SU(3)_c\times U(1)_{\text{em}}$ is achieved by the doublet  vevs of $\utilde{\bf 5}_S\oplus\utilde{\overline{\bf 5}}_S$. The two last terms in the first line of Eq.(\ref{eq:SU5W}) generate an effective $\mu$-term $\mu_{\text{eff}}=\mu-\tfrac{3m\beta}{\lambda}$ for $H_{U,D}$ (embedded in $\utilde{\bf 5}_S\oplus\utilde{\overline{\bf 5}}_S$), while the supersymmetric mass associated to the colour-triplets $H_3,\overline{H}_3$ of $\utilde{\bf 5}_S\oplus\utilde{\overline{\bf 5}}_S$ is $\mu+\tfrac{2m\beta}{\lambda}$. From the phenomenological perspective, it is necessary to impose the fine-tuning $|\mu_{\text{eff}}|\ll|\mu|,|\tfrac{3m\beta}{\lambda}|$, making the doublets light and the colour-triplets superheavy: while this requirement is non-natural, it remains technically natural once set, due to the protection by
		supersymmetry --- contrarily to the situation in non-supersymmetric $SU(5)$. Further model-building ingredients can also naturally protect the doublet mass term.
		
		The terms in the second line of Eq.(\ref{eq:SU5W}) are at the origin of the Yukawa couplings in the MSSM. Explicit decomposition immediately provides that $Y_5$ generates Yukawa couplings for the down-type quarks and leptons while $Y_{10}$ gives the up-type Yukawa. For the third generation, the resulting unification of down and lepton masses at the GUT scale ($m^{\text{GUT}}_{b}=m^{\text{GUT}}_{\tau}$) roughly yields $m_b\sim3m_{\tau}$ at low energy, which is in acceptable agreement with the measurements. This does not work for the lighter generations. Noting however that $\utilde{\overline{\bf 5}}\otimes\utilde{\bf 10}=\utilde{\bf 5}\oplus\utilde{\bf 45}$ and
		$\utilde{\bf 10}\otimes\utilde{\bf 10}=\utilde{\overline{\bf 5}}\oplus\utilde{\overline{\bf 45}}\oplus\utilde{\bf 50}$, one can introduce additional `Higgs' superfields in a $\utilde{\bf 45}_S\utilde{\oplus\overline{\bf 45}}_S$ representation, which allow for further terms of Yukawa type and contain doublets accepting a
		vev. The Clebsch-Gordan coefficients at $M_{\text{GUT}}$ then produce $m^{\text{GUT}}_{\mu}=3m^{\text{GUT}}_s$ (for the second generation; in the absence of contributions from the
		$\utilde{\bf 5}_S\oplus\utilde{\overline{\bf 5}}_S$ terms), resulting in the phenomenologically acceptable $m_{\mu}\sim m_s$. Thus, models involving both $\utilde{\bf 5}_S\oplus\utilde{\overline{\bf 5}}_S$ and $\utilde{\bf 45}_S\oplus\utilde{\overline{\bf 45}}_S$ are potentially viable explanations of the unification of lepton and down-type masses. Similarly, the $SU(5)$ symmetry implies the unification (at $M_{\text{GUT}}$) of numerous soft terms, such as the quadratic terms for all fields belonging to the same representation. As long as
		supersymmetry-breaking is communicated to the observable sector {\it via} an $SU(5)$ singlet, the gaugino masses also unify (at $M_{\text{GUT}}$) $M^{\text{GUT}}_1=M^{\text{GUT}}_2=M^{\text{GUT}}_{3}$, leading to the popular hierarchy $M_3\approx3M_2\approx6M_1$ at low energy.
		
		Beyond $SU(5)$, the immediate benefit of considering unification groups of higher rank, such as $SO(10)$ or $E_6$, rests with the automatic cancellation of chiral anomalies. The branching rules of $SO(10) \supset SU(5)\times U(1)$:\index{GUT!$SO(10)$}
		\begin{align}
			\utilde{\bf 16}&=(\utilde{\overline{\bf 5}},-3)\oplus(\utilde{\bf 10},1)\oplus(\utilde{\bf 1},-5)\,,
			\hspace{1cm}\utilde{\bf 10}=(\utilde{\bf 5},2)\oplus(\utilde{\overline{\bf 5}},-2)\,,\nonumber\\
			\utilde{\bf 45}&=(\utilde{\bf 24},0)\oplus(\utilde{\bf 10},4)\oplus(\utilde{\overline{\bf 10}},-4)\oplus(\utilde{\bf 1},0)\\
			\utilde{\bf 120}&=(\utilde{\bf 5},2)\oplus(\utilde{\overline{\bf 5}},-2)\oplus(\utilde{\bf 10},-6)\oplus(\utilde{\bf 45},2)\oplus(\utilde{\overline{\bf 45}},-2)\nonumber\\
			\utilde{\bf 126}&=(\utilde{\bf 1},-10)\oplus(\utilde{\overline{\bf 5}},-2)\oplus(\utilde{\bf 10},-6)\oplus(\utilde{\overline{\bf 15}},6)\oplus(\utilde{\bf 45},2)\oplus(\utilde{\overline{\bf 50}},-2)\nonumber
		\end{align}
		suggest an embedding of all quark and lepton superfields within three generations of $\utilde{\bf 16}_F^f$, also containing $SU(5)$ singlets amounting to right-handed neutrinos. Then, observing that $\utilde{\bf 16}\otimes\utilde{\bf 16}=
		\utilde{\bf 10}\oplus\utilde{\bf 120}\oplus\utilde{\bf 126}$, one can write Yukawa couplings employing Higgs fields in the real
		$\utilde{\bf 10}_S$, $\utilde{\bf 120}_S$ or the complex $\utilde{\bf 126}_S\oplus\utilde{\overline{\bf 126}}_S$ representations. This leaves ample maneuvering space to accommodate the fermion masses, which a single Higgs representation would fail to explain due to phenomenologically unphysical Yukawa unifications. In particular, in order to generate a `Majorana mass term' for the right-handed neutrinos and allow for a Type-I seesaw, the residual $U(1)$ symmetry needs to be broken. This can be achieved with a $\utilde{\bf 16}_S$ or a $\utilde{\bf 126}_S$  vev, with the difference that the second choice preserves a residual $\mathbb Z_2$-symmetry, which can be understood as $R$-parity. The $\utilde{\bf 5}_S\oplus\utilde{\overline{\bf 5}}_S$ Higgs fields of $SU(5)$ are embedded within a $\utilde{\bf 10}_S$ representation of $SO(10)$. Finally, $SU(5)$ can be broken with the $\utilde{\bf 24}_S$ contained within the $\utilde{\bf 45}_S$ representation of $SO(10)$ (or a $\utilde{\bf 54}$ or a $\utilde{\bf 210}$). Soft-breaking terms {\it a priori} unify more completely at $M_{\text{GUT}}$ than in $SU(5)$, since all MSSM fields are collected within $\utilde{\bf 16}_F^f$ and $\utilde{\bf 10}_S$. An even more ambitious unification pattern is possible in $E_6$, with both $\utilde{\bf 16}_F^f$ and $\utilde{\bf 10}_S$ of $SO(10)$ collected within a $\utilde{\bf 27}$ of $E_6$, which we will not discuss here.
		
		Matter stability is usually presented as the weak point of grand unified theories. Indeed, the interactions mediated by the $(\utilde{\bf 3}, \utilde{\bf 2}, -\tfrac{5}{6})\oplus(\utilde{\bf \bar{3}}, \utilde{\bf 2}, \tfrac{5}{6})$ gauge bosons contribute to effective dimension $6$ baryon- and lepton-number violating low-energy operators, which can mediate proton decay, {\it via} channels such as $p\to e^+\pi^0$. The typical associated lifetime scales like $\tau_p\sim M_{\text{GUT}}^4/(\alpha_U^2M_P^5)$,
		where $M_P$ is the proton mass  and $\alpha_U\equiv g_U^2/(4\pi)$ with $g_U$ the gauge coupling at the unification scale. For a non-supersymmetric Grand Unification, the predicted lifetime tends to be incompatible with the corresponding experimental limit. In a supersymmetric context, the larger $M_{\text{GUT}}$ apparently provides some breathing space. However, the existence of comparatively light scalar partners to the standard fermions alters the validity of the naive analysis, since it allows the emergence at $M_{\text{soft}}$ of baryon- and lepton-number operators of dimension $5$ involving such fields, which in turn, below $M_{\text{soft}}$, alter the $M_{\text{GUT}}^{-2}$ suppression of the contributions to dimension $6$ operators mediating proton decay. Due to the structure of these operators, the proton decay modes that they funnel typically involve matter of higher-generation content, such as kaons, muons, muon and $\tau$ neutrinos. In practice, a loop suppression factor alleviates the contributions {\it via} such intermediate operators of dimension $5$, so that these do not necessarily ruin the viability of supersymmetric Grand Unification: a detailed analysis is needed in each case. In fact, purely scalar dimension $4$ operators at $M_{\text{soft}}$ can also mediate proton decay, but contribute only at two-loop. In all this discussion, we have assumed the existence of a matter parity, which constrains baryon- and lepton-number violating effects to appear in operators of higher dimensions (or involving scalar partners). The pattern of violation is modified if $R$-parity is not satisfied at $M_{\text{soft}}$, which we consider in the following section.
		
		\subsection{\boldmath $R$-parity-violating phenomenology}
		\index{R-parity}
		So far, we have forbidden the terms in the second line of Eq.(\ref{eq:MSSMW}), and their soft-breaking counterparts, by requesting the model to satisfy $R$-parity. The original motivation behind this choice was to forbid the emergence of baryon- (in the last term of this equation: `$UDD$') or lepton-number violating effects (in the three previous terms: `$LH$', `$LLE$' and `$LQD$'). As we noticed in the discussion concerning Grand Unification, however, $R$-parity does not forbid the emergence of non-renormalisable operators violating baryon- and lepton-number, as one would expect if the MSSM is only a low-energy effective field theory of some higher-energy dynamics; thus it does not fully protects matter stability. We may therefore relax the assumption of $R$-parity conservation altogether, which results in significant alterations of the phenomenology. We refer the reader to Ref.\cite{R-pa}.
		
		Baryon-number violating processes, such as nucleon decay, obviously represent the first challenge to $R$-parity violation, as new physics at $M_{\text{soft}}$ potentially contributes to such phenomena. Here, we stress that proton decay usually requires both lepton- and baryon-number violation, because leptons (electron, muon, neutrinos) are the only available fermions below the proton mass (in the absence of exotic particles) and that the angular momentum should be preserved in the transition. New discrete symmetries, such as baryon triality, lepton triality or proton hexality, can be invoked to forbid nucleon decays. Pure baryon-violating processes such as di-nucleon decay or neutron-antineutron oscillations place further limits on couplings of $UDD$-type, but they usually depend on the pattern of flavour violation as well, due to the antisymmetry of the baryon-number violating $UDD$ operators with respect to down-quark generation indices. In the presence of a light exotic state, usually a bino, below the proton mass, proton decay can occur without lepton-number violation. The baryon-number violating couplings thus appear as more immediately constrained and are often explicitly eliminated by the call to a symmetry.
		
		As we implied above, the bino state may be light if it is unstable, which is the case if $R$-parity is violated: cosmological and astrophysical bounds are indeed no longer applicable. If the bino is long-lived, its decays might be observed in far detectors, as currently deployed at the Large Hadron Collider. More generally, collider phenomenology may significantly change in the presence of $R$-parity violating effects, as supersymmetric particles can now be produced as single resonances while the missing energy signature is no longer guaranteed. Searches with single leptons, jets and little missing energy may then prove viable alternative strategies. The lightest supersymmetric particle may still be a Dark Matter candidate if it is sufficiently long-lived to persist on cosmological timescales. In such a case, it needs to be neutral and satisfy conditions relative to its abundance at the time of recombination. Alternatively, a charged or coloured particle with short lifetime is eligible as lightest supersymmetric particle for collider physics, without implications for cosmology.
		
		Given that $R$-parity is no longer present to distinguish between standard fields and their supersymmetric partners, the latter usually mix, such as leptons with uncoloured gauginos and higgsinos, or sleptons with Higgs bosons. This mixing should generally remain subleading so that standard particles retain their usual properties. It also entails some degree of ambiguity in the definition of the superfields: a convenient, though not mandatory choice consists in requesting that the sneutrino fields do not take vevs. In addition, the emergence of neutrino masses is an interesting consequence of the mixing of neutrinos with neutral gauginos and higgsinos. At tree level, only one mass, scaling like $(\mu_i M_Z/\mu)^2/M_2$ develops; comparison with observed limits place bounds on $|\mu_i/\mu|,|\mu_i/M_2|$ of the order $10^{-5}$. Loop corrections involving couplings of $LLE$ or $LQD$ type produce a second neutrino mass, opening the possibility for an explanation of neutrino oscillations in terms of pure $R$-parity violating effects.
		
		The {\it a priori} independent flavour- and CP-violating pattern of the trilinear $R$-parity violating couplings can be constrained in low-energy observables. Such couplings indeed contribute at tree-level (typically {\it via} a sfermion exchange) to precisely known observables, such as lepton and meson decays or particle-antiparticle oscillations, quark- or lepton-flavour transitions, the unitarity triangle, the anomalous magnetic dipole moments of leptons, the electric dipole moments of leptons and atoms, neutrinoless double beta decay, {\it etc.} A comprehensive coverage of these bounds is usually performed under simplifying assumptions, such as the dominance of a single $R$-parity violating coupling, or a pair of couplings, or in numerical form, as the diversity of available terms is otherwise difficult to efficiently  tackle.
		
		From the perspective of Grand Unification, we have  mentioned how the conservation / violation of $R$-parity could be related to $SO(10)$ breaking. In $SU(5)$, the $R$-parity violating superpotential summarizes to: \index{R-parity}
		\begin{equation}
			W_{SU(5)}^{\text{RpV}}=\kappa_f(\utilde{\overline{\bf 5}}_F^f)^i(\utilde{\bf 5}_S)_i+\frac{1}{2}\lambda^{SU(5)}_{fgh}(\utilde{\overline{\bf 5}}_F^f)^i(\utilde{\overline{\bf 5}}_F^g)^j(\utilde{\bf 10}_F^h)_{ij}\,.
		\end{equation}
		In this case, trilinear couplings are unified and the limits from proton decay makes such couplings negligible at the electroweak scale. The coupling constants $\kappa_i$  are also phenomenologically very small due to  their  implications for neutrino physics.
		
		\smallskip	
		As a summary of this phenomenological overview of supersymmetry / supergravity in particle physics, we have seen that very diverse types of effects could emerge within such theories, due to the rich field content and the {\it a priori} uncontrolled pattern of couplings inherited from their construction. Such models thus raise considerable challenge to model building for `natural' explanations of hierarchies, although these are now stabilised by supersymmetry ({\it i.e.,}~technically natural). The most simplistic model-building assumptions are usually incompatible with the experimental situation, leading to an increase in complexity in the field content and the parameter space, where one had originally wished a predictive and symmetry-constrained structure. Conversely, this diversity makes supersymmetry-inspired models a good laboratory to study and motivate non-standard effects in particle physics.

                \bibliographystyle{spbasic}
		\bibliography{ref}
		
\printindex		
	\end{document}